\documentclass[aps, prb, twocolumn, superscriptaddress]{revtex4-1}

\usepackage[]{amsmath}
\usepackage{amssymb}

\usepackage{fontawesome}

\usepackage[]{graphicx}
\usepackage[]{mathtools}
\usepackage[]{bm}
\usepackage[caption=false]{subfig}

\newcommand{\figref}[1]{\figurename~\ref{#1}}

\usepackage{cases}

\usepackage[]{comment}

\begin{document}

\title{Optically induced electron spin currents in the Kretschmann configuration}

\author{Daigo Oue}
\affiliation{Kavli Institute for Theoretical Sciences, University of Chinese Academy of Sciences, Beijing, 100190, China.}
\affiliation{The Blackett Laboratory, Department of Physics, Imperial College London, Prince Consort Road, Kensington, London SW7 2AZ, United Kingdom}
\author{Mamoru Matsuo}
\affiliation{Kavli Institute for Theoretical Sciences, University of Chinese Academy of Sciences, Beijing, 100190, China.}
\affiliation{CAS Center for Excellence in Topological Quantum Computation, University of Chinese Academy of Sciences, Beijing 100190, China}

\date{\today}

\begin{abstract}
  We investigate electron spin currents induced optically via plasmonic modes in the Kretschmann configuration.
  By utilising the scattering matrix formalism,
  we take the plasmonic mode coupled to external laser drive into consideration and calculate induced magnetization in the metal.
  The spatial distribution of the plasmonic mode is inherited by the induced magnetization,
  which acts as inhomogeneous effective magnetic field and causes the Stern-Gerlach effect to drive electron spin currents in the metal.
  We solve spin diffusion equation with a source term to analyze the spin current as a function of the spin diffusion length of the metal, the frequency and the incident angle of the external drive.
\end{abstract}

\maketitle

\section{Introduction}
\label{sec:introduction}
Electron spin is one of the fundamental physical quantities, which can carries heat, angular momentum, and even quantum information \cite{flipse2014observation,daimon2016thermal,zolfagharkhani2008nanomechanical,harii2019spin,vrijen2000electron,wesenberg2009quantum},
and thus it is crucial to control the electron spin by other excitations.
One way to manipulate the electron spin is to use inhomogeneous magnetic field,
where Stern-Gerlach-type force drives spin currents carried by conduction electrons \cite{fabian2002spin,vzutic2002spin,fabian2002theory}.
Recently, it has been reported that circulation or non-zero vorticity of electric charge flow induce spin current in the metal \cite{takahashi2016spin, kobayashi2017spin, okano2019nonreciprocal}.
In these studies, the non-zero vorticity is effective inhomogeneous magnetic field which also brings about the Stern-Gerlach-type effect and drives the electron spin transport.
Keeping these ideas in mind,
we consider optically driving electron spin current via plasmonic modes in the Kretschmann configuration \cite{maier2007plasmonics}.

\par
Recently, the spin current generation from light via plasmonic modes have been studied \cite{uchida2015generation, ishii2017wavelength, oue2020electron, oue2020effects}.
In Refs. \cite{uchida2015generation, ishii2017wavelength}, 
they investigated spin current generation in magneto-plasmonic systems at localised plasmon resonance conditions,
where the difference of the nonequilibrium distribution functions between magnons and plasmons is utilized as in the spin pumping and the spin Seebeck effect \cite{matsuo2018spin, kato2019microscopic}.
In our previous works \cite{oue2020electron, oue2020effects},
we proposed the angular momentum conversion from the transverse spin of propagating surface plasmons (SPs) to electronic systems,
where the transverse spin circulates electron gas to induce inhomogeneous magnetic field and result in electron spin current in the metallic media.

\par
However, to the best of our knowledge, 
it has never theoretically investigated the electron spin dynamics in a plasmonic system where the plasmonic mode is coupled to external laser drive.
In this study, we calculate the electron spin current induced optically in a trilayer system.

\par
\begin{figure}[htbp]
  \centering
  \includegraphics[width=.8\linewidth]{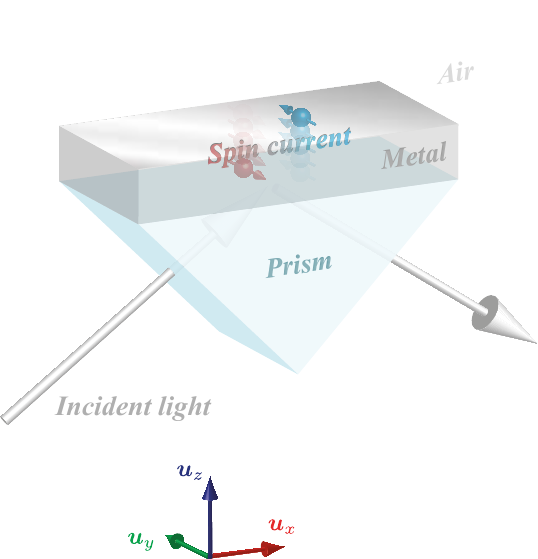}
  \caption{
    Optically driven electron spin current in a trilayer system (Kretschmann configuration),
    where a metal film ($\epsilon = \epsilon^m$) with a thickness of $d$ is sandwiched by two dielectrics, air ($\epsilon^a = 1$) and prism ($\epsilon^p = 1.5^2$).
    A laser drive field is incident on the metal from the prism side,
    and surface plasmon modes can be excited on the air-metal interface when the incident angle exceeds the critical angle ($\theta_\mathrm{in} > \theta_\mathrm{c} \equiv \arcsin \sqrt{\epsilon^a/\epsilon^p}$).
    In the excited plasmonic modes, the electric field is circulating,
    which induces orbiting motion of electron gas in the metal and results in inhomogeneous magnetization.
    The electron spin experiences the Stern-Gerlach effect under the magnetization,
    and the electron spin current is driven in the direction perpendicular to the air-metal interface.
  }
  \label{fig:fig1}
\end{figure}

We consider a metal film sandwiched by two different dielectrics (\figref{fig:fig1}).
Let the permittivity of the metal be $\epsilon^m$ and those of dielectrics $\epsilon^a$ and $\epsilon^p$ ($\epsilon^a < \epsilon^p$).
This layered system is called the Kretschmann configuration which is one of the typical systems to excite plasmonic modes by a laser drive \cite{maier2007plasmonics}.
When the laser radiation is incident on the metal film from one dielectric side where the permittivity is higher than the other dielectric side,
surface plasmon modes can be excited above the critical angle condition ($\theta_\mathrm{in} > \theta_\mathrm{c} \equiv \arcsin (\sqrt{\epsilon^a/\epsilon^p})$).

\par
The excited plasmonic modes have exponential profile and thus transversally spinning electric field in the metal due to the spin-momentum locking effect of light in nonparaxial regime \cite{bliokh2012transverse, bliokh2014extraordinary, bekshaev2015transverse, bliokh2015quantum, van2016universal, fang2017intrinsic, oue2019dissipation}.
When the frequency of the field is smaller than the plasma frequency of the metal,
the highly confined spinning field induces the electron gas to circulate,
which generates the steep gradient of magnetization.
Although the magnetization itself is so small that it is even difficult to detect as analyzed in the previous studies \cite{bliokh2017optical, bliokh2017optical_new_j_phys},
its gradient can be large enough to drive electron spin transport \cite{oue2020electron, oue2020effects}.

\section{Magnetization induced by plasmonic fields}
\label{sec:magnetication_induced_by_plasmonic_fields}
To investigate the electron dynamics in the plasmonic field,
we firstly calculate the electromagnetic field by a scattering matrix formalism,
and evaluate magnetization induced by the field in the metal.
Once we obtain the frequency dependence, the incident angle dependence, and the spatial distribution of the magnetization,
the diffusive dynamics of electron spin can be calculated from spin diffusion equation with a source term \cite{matsuo2013mechanical,matsuo2017theory}.

\subsection{Scattering matrix formalism for a multilayer system}
\label{subsec:scattering_matrix_formalism_for_a_multilayer_system}
\begin{figure}[htbp]
  \centering
  \includegraphics[width=.8\linewidth]{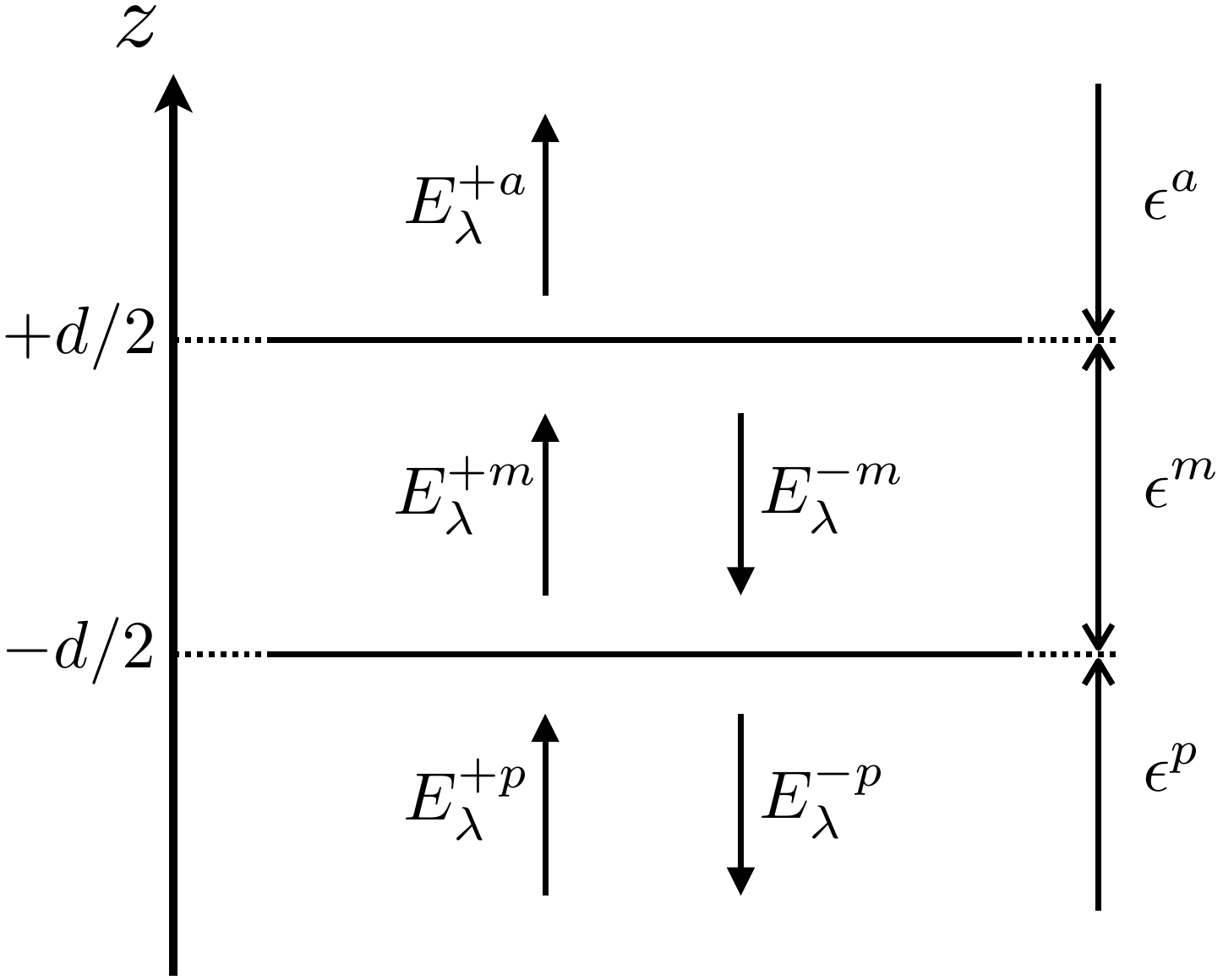}
  \caption{
    Electromagnetic field in each layer is expanded by the eigenmodes as in \eqref{eq:E} and \eqref{eq:H}.
    We set the amplitudes of the fields with a polarization of $\lambda=s,p$ $E_{\lambda}^{\sigma\tau}$,
    where $\sigma=\pm$ and $\tau=a,m,p$ specify the direction of the $z$ components of the wavevectors and the medium,
    respectively (i.e., the permittivity in each layer is $\epsilon^{a,m,p}$).
    The frequency $\omega$ and the wavenumber in the direction parallel to the interfaces $\bm{k}_\parallel\ (|\bm{k}_\parallel|^2<\omega^2\epsilon^p/c^2)$ are free parameters.
    We impose the field continuity (boundary conditions) at $z=\pm d/2$ in order to derive simultaneous equations determining the scattering matrix of this trilayer system \eqref{eq:scattering_symb}.
  }
  \label{fig:EM_calc}
\end{figure}

We start from time-harmonic Maxwell equations in dielectrics,
\begin{subnumcases}{}
  \nabla \cdot \bm{E} = \nabla \cdot \bm{H} = 0,
  \label{eq:Gauss}
  \\
  \nabla \times \bm{E} - i\omega\mu_0 \bm{H} = 0,
  \label{eq:Faraday}
  \\
  \nabla \times \bm{H} + i\omega\epsilon^\tau\epsilon_0 \bm{E} = 0.
  \label{eq:Ampere}
\end{subnumcases}
Here, we have used the vacuum permittivity $\epsilon_0$ and the vacuum permeability $\mu_0$,
and the superscript $\tau=a,m,p$ specifies the permittivity in each layer (see \figref{fig:EM_calc}).

\par
Our trilayer system has translational invariance in $x$ and $y$ direction,
and it is convenient to use wavenumber, $k_x$ and $k_y$, as parameters rather than $x$ and $y$,
Also, we work in the frequency domain (i.e., use $\omega$ rather than $t$).
By using Fourier integral,
our electric field can be written as
\begin{align}
  \bm{E} = \int \frac{d\bm{k}_\parallel}{(2\pi)^2} \int \frac{d\omega}{2\pi}\ 
  \bm{E}(\bm{k}_\parallel,\omega) e^{i\bm{k}_\parallel \cdot \bm{r}_\parallel - i\omega t},
  \label{eq:E_fourier}
\end{align}
and the Maxwell equations lead to a wave equation,
\begin{align}
  \left[
    -\frac{\partial^2}{\partial z^2} - \left(\epsilon^\tau {k_0}^2 - {k_\parallel}^2\right)
  \right]
  \bm{E}(\bm{k}_\parallel, \omega) = 0.
  \label{eq:wave_eq}
\end{align}
Here, we have used parallel wavenumber and the radiation wavenumber in vacuum,
\begin{align*}
  \bm{k}_\parallel 
    &= k_x \bm{u}_x + k_y \bm{u}_y,
    \quad
    k_\parallel = \sqrt{{k_x}^2+{k_y}^2},
    \\
  k_0 &= \frac{\omega}{c}.
\end{align*}
We define the unit vectors $\bm{u}_{x,y,z}$ along corresponding axes as indicated in \figref{fig:fig1}.
We can easily solve Eq. \eqref{eq:wave_eq} in each layer to get the electric field,
\begin{align}
  \bm{E}(\bm{k}_\parallel,\omega) &= \sum_{\sigma=\pm} \bm{C}^{\sigma \tau} e^{i\sigma K^\tau z},
\end{align}
where $\sigma=\pm$ specifies the propagation direction (i.e., $\sigma=\pm$ means that the wave propagates in the $\pm z$ direction),
and the wavenumber in the $z$ direction in each layer $l$ is defined by
\begin{align}
  K^\tau(k_\parallel, \omega) &= \sqrt{\epsilon^\tau {k_0}^2 - k_\parallel^2}.
  \label{eq:wavevector}
\end{align}
The coefficient vector $\bm{C}^{\sigma \tau}$ is fixed by Eq. \eqref{eq:Gauss}, 
\begin{align}
  \nabla \cdot \bm{E} &= 0
  \notag \\
  \sum_{\sigma} \bm{k}^{\sigma \tau} \cdot \bm{C}^{\sigma \tau} e^{i\sigma K^\tau z} &= 0,
  \label{eq:Gauss_alt}
\end{align}
where we have defined
\begin{align*}
  \bm{k}^{\sigma\tau} (k_\parallel,\omega) &= \bm{k}_{\parallel} + \sigma K^\tau(k_\parallel,\omega) \bm{u}_z.
\end{align*}
Since Eq. \eqref{eq:Gauss_alt} has to be satisfied for arbitrary $z$,
the coefficient vector $\bm{C}^{\sigma \tau}$ should be orthogonal to the corresponding wavevector $\bm{k}^{\sigma \tau}$.
There are two candidate vectors given by vector products,
\begin{align}
  \bm{e}_{\lambda}^{\sigma \tau}(\bm{k}_\parallel,\omega)
  &=
  \begin{cases}
    \cfrac{\bm{k}^{\sigma \tau} \times \bm{u}_z}
    {|\bm{k}^{\sigma \tau} \times \bm{u}_z|} 
    & \lambda = s,\\
    \cfrac{\bm{k}^{\sigma \tau} \times \bm{k}^{\sigma \tau}\times \bm{u}_z}
    {|\bm{k}^{\sigma \tau} \times \bm{k}^{\sigma \tau} \times \bm{u}_z|} 
    & \lambda = p,
  \end{cases}
  \label{eq:polarization_vector}
\end{align}
where we have labeled the two vectors by $\lambda=s, p$.
Thus, we can set
\begin{align}
  \bm{C}^{\sigma \tau} = \sum_{\lambda=s,p} E_{\lambda}^{\sigma \tau} \bm{e}_{\lambda}^{\sigma\tau},
\end{align}
where $E_{\lambda}^{\sigma \tau}$ is the modal amplitude of the field with given polarization and propagation direction in each layer.
Finally, we have
\begin{align}
  \bm{E}(\bm{k}_\parallel,\omega) 
  &= \sum_{\lambda\sigma} E_{\lambda}^{\sigma\tau} \bm{e}_{\lambda}^{\sigma\tau} e^{i\sigma K^\tau z}.
  \label{eq:E}
\end{align}

\par
In order to obtain the corresponding magnetic field,
we substitute Eq. \eqref{eq:E} into \eqref{eq:Faraday},
\begin{align}
  \bm{H}(\bm{k}_\parallel,\omega)
  &= \frac{1}{\omega \mu_0}
  \sum_{\sigma=\pm}
  \bm{k}^{\sigma\tau}
  \times
  \bm{E}(\bm{k}_\parallel,\omega),
  \notag
  \\
  &= \frac{1}{Z_0} \sum_{\lambda \sigma}
  E_{\lambda}^{\sigma \tau}\operatorname{sgn}(\lambda)
  \sqrt{\epsilon^\tau} 
  \bm{e}_{\bar{\lambda}}^{\sigma \tau} e^{i\sigma K^\tau z}.
\label{eq:H}
\end{align}
where $Z_0=\sqrt{\mu_0/\epsilon_0}$ is the impedance of free space.
Note that we here have defined 
\begin{align*}
  \operatorname{sgn}(\lambda)=
  \begin{cases}
    +1 & \lambda=s,\\
    -1 & \lambda=p,
  \end{cases}
\end{align*}
$\bar{p} = s$ and $\bar{s} = p$.

Let us take the coordinate so that the wavevector lies in the $xz$ plane ($k_x = k, k_y = 0$),
and fix $\lambda = p$.
This is because we are interested in plasmonic excitations, and there is no SPs excited by the $s$ polarization.
Then, we have
\begin{align}
  \bm{e}_{p}^{\sigma\tau} &= 
  \frac{1}{\sqrt{\epsilon^\tau} k_0}
  \begin{pmatrix}
    \sigma K^\tau\\
    0\\
    -k
  \end{pmatrix},
  \quad
  \bm{e}_{s}^{\sigma\tau} =
  \begin{pmatrix}
    0\\
    -1\\
    0
  \end{pmatrix}.
  \label{eq:polarization_vector_in}
\end{align}

\par
Now, we are ready to derive the scattering matrix of our system.
Imposing the continuities of the tangential fields, $E_z$ and $H_y$, at the two interfaces ($z=\pm d/2$),
we can obtain simultaneous equations in the matrix form,
\begin{align}
  S^{-1} \psi_\mathrm{out} &= \psi_\mathrm{in}
  \label{eq:scattering_symb}
\end{align}
where the scattering matrix of our system is defined by
\begin{widetext}
\begin{align}
  S^{-1} &=
    \begin{pmatrix}
    -\cfrac{K^a}{\sqrt{\epsilon^a}} e^{+iK^ad/2} & -\cfrac{K^m}{\sqrt{\epsilon^m}}e^{-iK^md/2} & \cfrac{K^m}{\sqrt{\epsilon^m}} e^{+iK^md/2} & 0\\
    \sqrt{\epsilon^a}e^{+iK^ad/2} & -\sqrt{\epsilon^m} e^{-iK^md/2} & -\sqrt{\epsilon^m} e^{+iK^md/2} & 0\\
    0 & -\cfrac{K^m}{\sqrt{\epsilon^m}} e^{+iK^md/2} & \cfrac{K^m}{\sqrt{\epsilon^m}} e^{-iK^md/2} & \cfrac{K^p}{\sqrt{\epsilon^p}}e^{+iK^pd/2}\\
    0 & \sqrt{\epsilon^m} e^{+iK^md/2} & \sqrt{\epsilon^m} e^{-iK^md/2} & -\sqrt{\epsilon^p} e^{+iK^pd/2}
  \end{pmatrix},
  \label{eq:S_inv}
\end{align}
\end{widetext}
and output and input vectors,
\begin{align}
  \psi_\mathrm{out} &= 
  \begin{pmatrix}
    E_{p+}^a\\
    E_{p-}^m\\
    E_{p+}^m\\
    E_{p-}^p
  \end{pmatrix},
  \quad
  \psi_\mathrm{in} =
  \begin{pmatrix}
    0\\
    \cfrac{K^p}{\sqrt{\epsilon^p}} e^{-iK^pd/2}\\
    0\\
    \sqrt{\epsilon^p} e^{-iK^pd/2}
  \end{pmatrix}
  E_{p+}^p.
  \label{eq:psi_out,in}
\end{align}
The $2 \times 2$ block matrices in the diagonal entries of $S^{-1}$ contain the information of plasmonic modes.
Indeed, the determinant of the left top block matrix returns the standard dispersion relation of the SP mode at the air-metal interface \cite{maier2007plasmonics},
\begin{align}
  \det
  \begin{pmatrix}
    -\cfrac{K^a}{\sqrt{\epsilon^a}} e^{+iK^ad/2} & -\cfrac{K^m}{\sqrt{\epsilon^m}} e^{-iK^md/2}\\
    \sqrt{\epsilon^a}e^{+iK^ad/2} & -\sqrt{\epsilon^m} e^{-iK^md/2}
  \end{pmatrix} = 0,\notag \\
  \cfrac{K^m/\epsilon^m}{K^a/\epsilon^a} = -1,
  \label{eq:dispersion_airmetal}
\end{align}
whereas that of the right bottom block matrix returns the dispersion of the SP mode at the metal-prism,
\begin{align*}
  \det
  \begin{pmatrix}
    \cfrac{K^m}{\sqrt{\epsilon^m}} e^{-iK^ad/2} & \cfrac{K^p}{\sqrt{\epsilon^p}} e^{+iK^pd/2}\\
    \sqrt{\epsilon^m}e^{-iK^md/2} & -\sqrt{\epsilon^p} e^{+iK^pd/2}
  \end{pmatrix} = 0,\\
  \cfrac{K^m/\epsilon^m}{K^p/\epsilon^p} = -1.
\end{align*}

\par
\begin{figure*}[htbp]
  \subfloat[$\theta_\mathrm{in}=30^\circ < \theta_\mathrm{c}$\label{subfig:Hy30deg}]{%
    \includegraphics[width=0.33\linewidth]{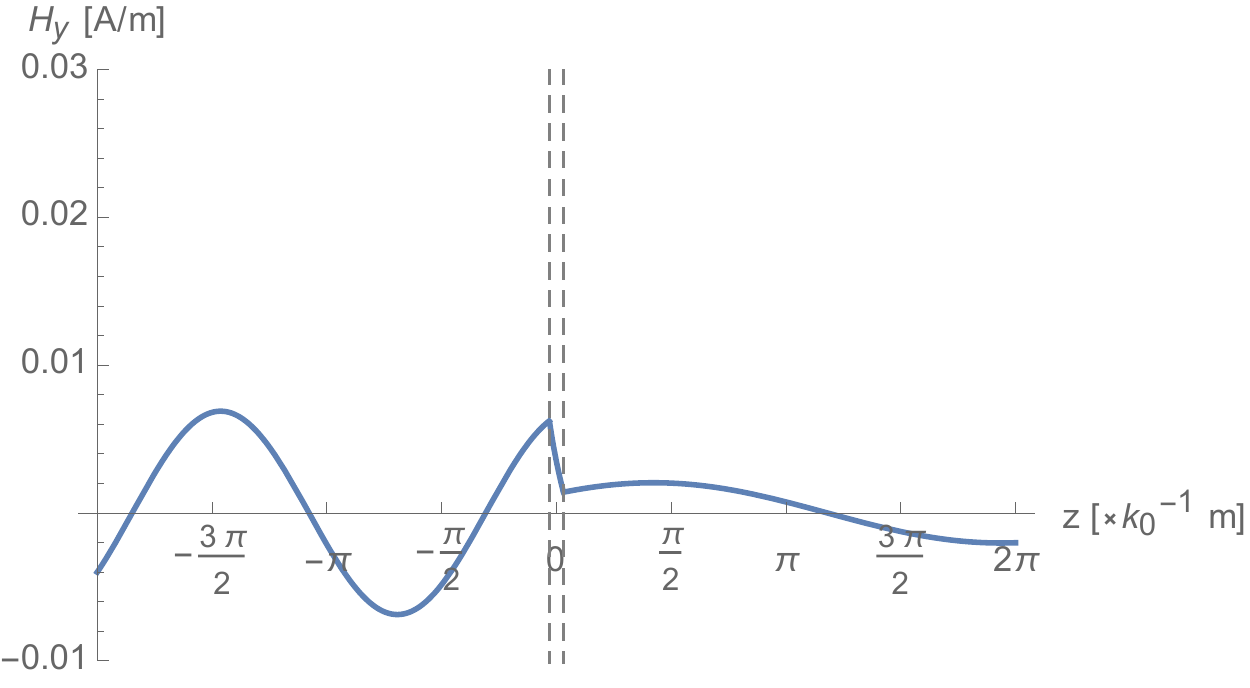}%
}\hfill
\subfloat[$\theta_\mathrm{in}=\theta_\mathrm{c}$\label{subfig:Hycritical}]{%
  \includegraphics[width=0.33\linewidth]{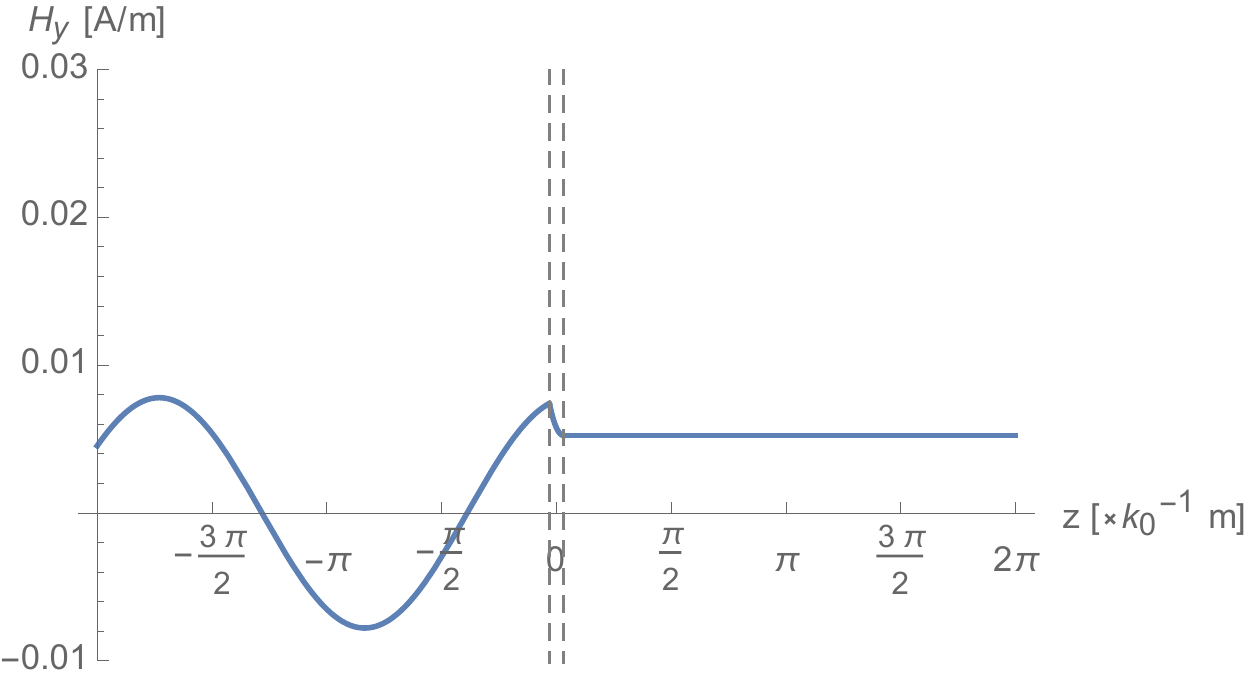}%
}\hfill
\subfloat[$\theta_\mathrm{in}=\theta_\mathrm{sp} > \theta_\mathrm{c}$\label{subfig:Hysp}]{%
  \includegraphics[width=0.33\linewidth]{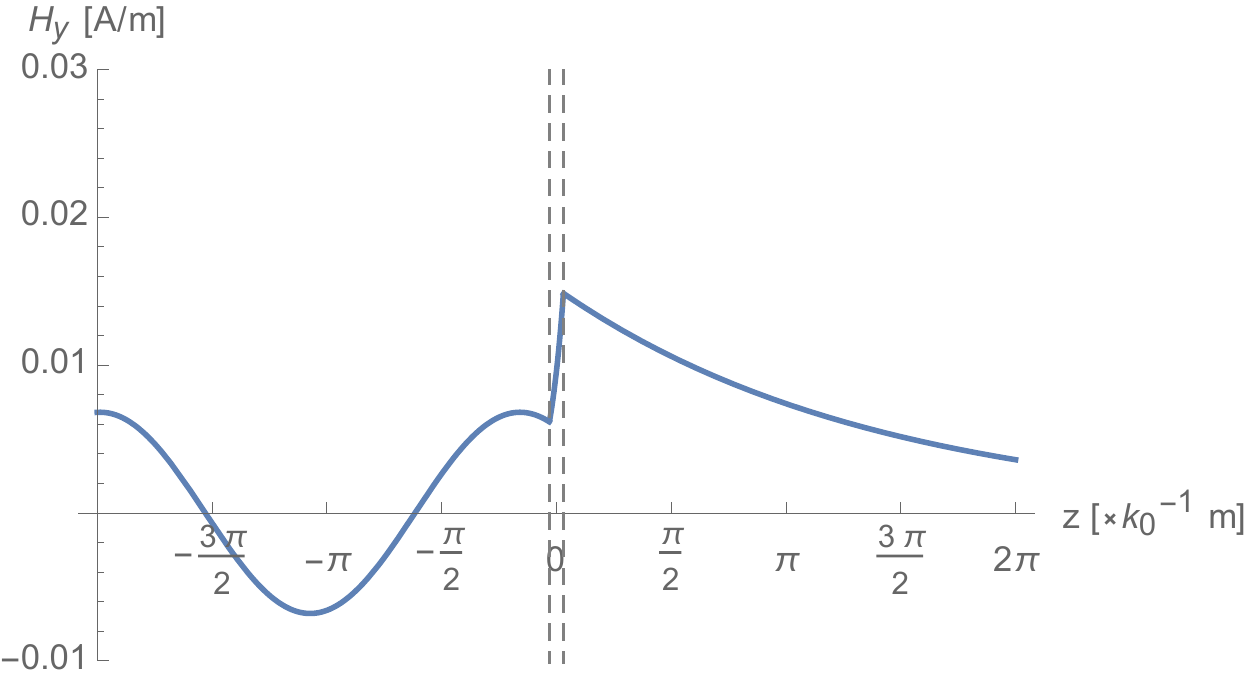}%
} 
  \caption{
    The field distribution in the trilayer system.
    The metal region ($-d/2 \leq z \leq d/2$) is indicated by the dashed grey lines.
    (a) The incident angle is smaller than the critical angle between the prism and the air layers.
    We can confirm that the incident field is decay in the metal layer and transmitted into the other side.
    (b) The critical angle of incidence $\theta_\mathrm{in}=\theta_\mathrm{c}$.
    At the critical angle, the electromagnetic field in the air layer is single plane wave, and the magnetic field distribution is uniform in the $z$ direction.
    (c) The incident angle is SP angle $\theta_\mathrm{sp}$ at which the $x$ component of the incident wavevector is matched with that of the SP.
    It is clear that the SP is excited at the interface between metal and air layers.
    The electromagnetic energy of the SP is compressed at the interface,
    and the field has exponential profile in the air region unlike the plane wave at the critical angle.
    In order to generate these plots,
    we set the incident frequency $\omega = c k_0 = 2.88 \times 10^{15}\ \mathrm{s^{-1}}\ (k_0=2\pi/655\ \mathrm{nm^{-1}})$,
    the incident amplitude $E_{p+}^p = 1.0\ \mathrm{V/m}$, 
    $d = 20\ \mathrm{nm}$,
    and $\omega_\mathrm{p} = 2\pi \times 2.1 \times 10^{15}\ \mathrm{s^{-1}}$ (the plasma frequency of gold).
  }
  \label{fig:distribution}
\end{figure*}

From \eqref{eq:scattering_symb}, we can derive the amplitudes of the electric and magnetic fields in the system excited by the given input (see \figref{fig:distribution}).
Here, we use the Drude free electron model for the permittivity of the metal,
\begin{align}
  \epsilon^m &= 1 - \frac{{\omega_\mathrm{p}}^2}{\omega^2}.
  \label{eq:Drude}
\end{align}

\par
We can define the incident angle such that
\begin{align}
  k &= \sqrt{\epsilon^p} k_0 \sin \theta_\mathrm{in},
  \quad
  K^p = \sqrt{\epsilon^p} k_0 \cos \theta_\mathrm{in}.
  \label{eq:theta_in}
\end{align}
In other words, the incident angle is given by
\begin{align}
  \theta_\mathrm{in} = \arctan \left(\frac{k}{K^p}\right).
\end{align}
When the incident angle is below the critical angle $\theta_\mathrm{c}$,
the incident field is transmitted into the other side as in the \figref{subfig:Hy30deg}.
In this case, there is no excited SP mode.
On the other hand,
if the incident angle exceeds the critical angle,
then the SP mode can be excited under wavenumber matching condition,
\begin{align}
  \sqrt{\epsilon^p} k_0 \sin \theta_\mathrm{in} &= k_\mathrm{sp}(\omega),
  \notag \\
  \theta_\mathrm{in} &= 
  \arcsin \left[
    \frac{k_\mathrm{sp}(\omega)}{\sqrt{\epsilon^p} k_0}
  \right]
  \equiv \theta_\mathrm{sp},
  \label{eq:wavenumber_matching}
\end{align}
where $k_\mathrm{sp}(\omega)$ can be determined by solving the dispersion relation \eqref{eq:dispersion_airmetal} for the wavenumber $k$.
In \figref{subfig:Hysp}, the SP mode excitation at the wavenumber matching condition \eqref{eq:wavenumber_matching} is shown.
While the field in the air layer has uniform spatial distribution in the critical incidence case ($\theta_\mathrm{in} = \theta_\mathrm{c}$) shown in \figref{subfig:Hycritical},
the field is compressed at the air-metal interface at the wavenumber matching condition.

\subsection{Inhomogeneous induced magnetization}
\label{subsec:inhomogeneous_induced_magnetization}
With the modal amplitudes calculated by the scattering matrix method in the previous part \ref{subsec:scattering_matrix_formalism_for_a_multilayer_system},
we can derive the angular momentum (AM) density of electron gas in the metal in terms of the electric field \cite{bliokh2017optical_new_j_phys, oue2020electron, oue2020effects},
\begin{align}
  \bm{S}_{el} 
  &= \frac{1}{4} \epsilon_0 \frac{d\epsilon}{d\omega} \mathfrak{Im}\left(\bm{E}^* \times \bm{E}\right) \\
  &= \frac{1}{2} \frac{k\mathfrak{Im}\left(K^m\right)}{\epsilon^m {k_0}^2}
  \frac{d\epsilon^m}{d\omega} \epsilon_0 \sum_\sigma \sigma |E_{p\sigma}^m|^2 e^{-2\sigma\mathfrak{Im}(K^m) z} \bm{u}_y
  \notag
  \\
  &\qquad (-d/2 \leq z \leq +d/2).
  \label{eq:S_el}
\end{align}
By simply multiplying the AM of the electron gas by the gyromagnetic ratio $\gamma=\frac{e}{2m}$,
we can evaluate the magnetization induced by the angular motion of the electron gas,
\begin{align}
  \bm{M} &= \frac{1}{2} \gamma \frac{k\mathfrak{Im}\left(K^m\right)}{\epsilon^m {k_0}^2}
         \frac{d\epsilon^m}{d\omega} \epsilon_0 \sum_\sigma \sigma |E_{p\sigma}^m|^2 e^{-2\sigma\mathfrak{Im}(K^m) z} \bm{u}_y.
\end{align}

\par
\begin{figure}[htbp]
  \centering
  \includegraphics[width=\linewidth]{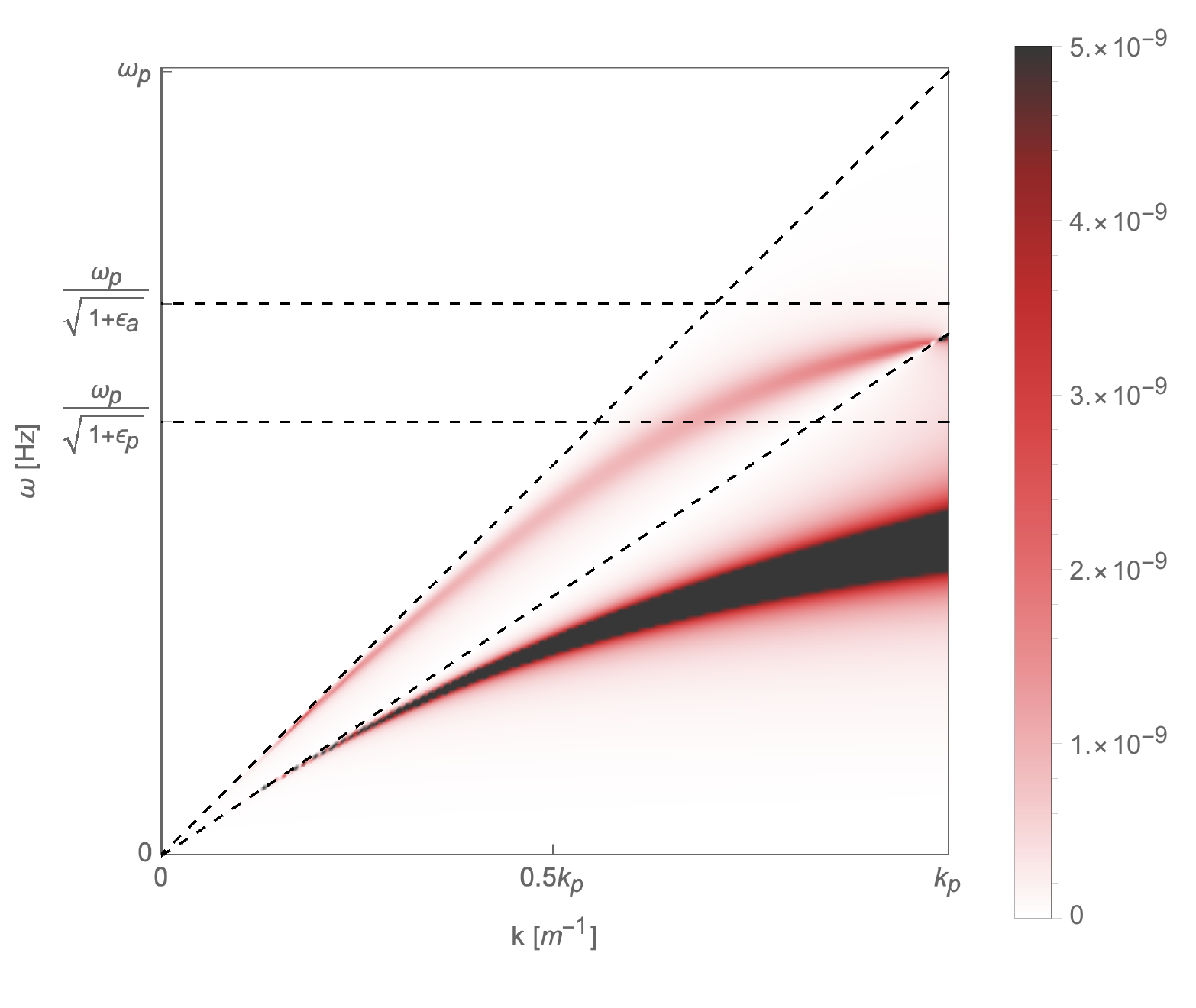}
  \caption{
    Magnetization at the metal surface $z=+d/2$ induced by the plasmonic modes as a function of the wavenumber in the $x$ direction and the frequency of the incident field.
    We can see two peaks in this colormap.
    This is because there are two types of plasmonic excitations in our system.
    One is the SP at the air-metal interface which mainly contributes to the upper peak.
    The other is the SP at the metal-prism interface, contributing to the lower peak.
    The light dispersion in the air and prism, $\omega=ck/\sqrt{\epsilon^a}$ and $\omega=ck/\sqrt{\epsilon^p}$, and the surface plasma frequencies at the two interfaces, $\omega_\mathrm{p}/\sqrt{1+\epsilon^a}$ and $\omega_\mathrm{p}/\sqrt{1+\epsilon^p}$, are indicated by dashed lines.
    In this figure, we set $E_{p+}^p = \sqrt{Z_0 \times 100\ \mathrm{mW/cm^2}} \approx 0.62 \times 10^3$ and $\omega_\mathrm{p}=2\pi \times 2.1 \times 10^{15}\ \mathrm{s^{-1}}$ (the plasma frequency of gold),
    and we defined $k_\mathrm{p} \equiv \omega_\mathrm{p}/c$.
    Note that we set the upper bound of this plot range at $5\ \mathrm{nA/m}$ because the peaks are so large that other data become invisible if we plot the full range.
  }
  \label{fig:magnetization}
\end{figure}
\begin{figure}[htbp]
  \centering
  \includegraphics[width=\linewidth]{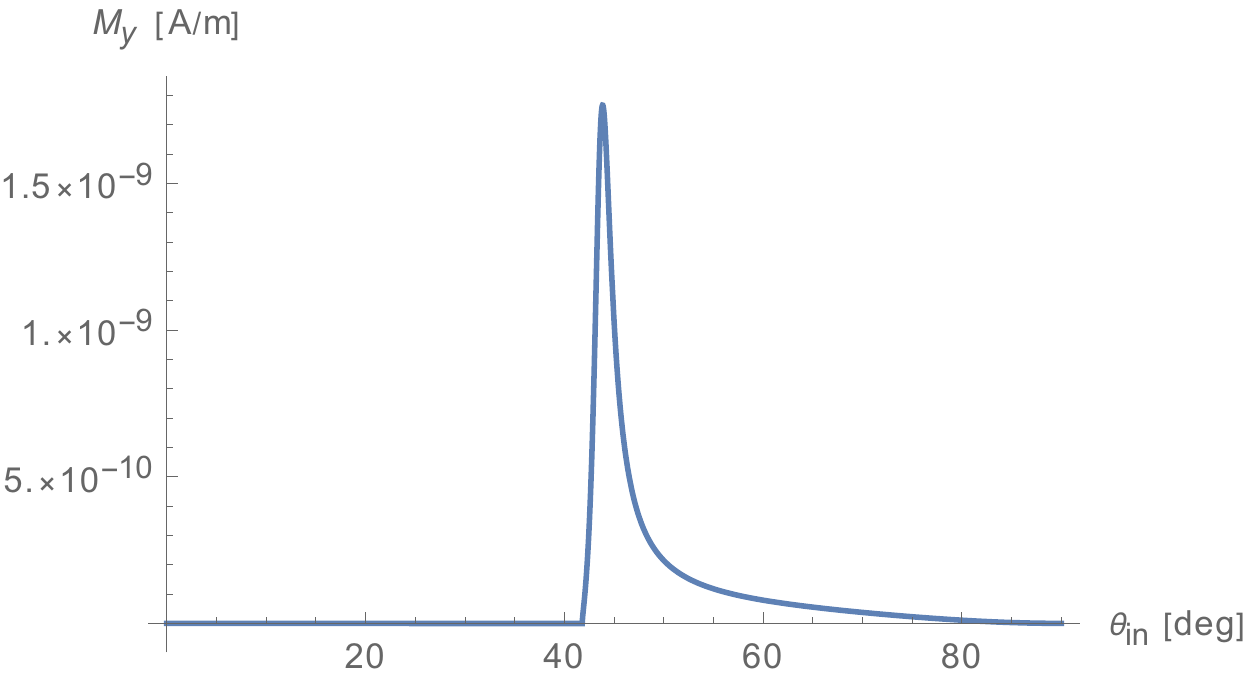}
  \caption{
    Induced magnetization at the metal surface $z=+d/2$ as a function of the incident angle.
    It is clear that it peaks at the SP angle of incidence ($\theta_\mathrm{in} = \theta_\mathrm{sp} \approx 43^\circ$).
    We set the incident frequency $\omega =c k_0 = 2.88 \times 10^{15}\ \mathrm{s^{-1}}\ (k_0=2\pi/655\ \mathrm{nm^{-1}})$, and all other parameters are the same as the previous figures.
  }
  \label{fig:magnetization_angle}
\end{figure}

In \figref{fig:magnetization}, the induced magnetization as a function of the incident field parameters, $\omega$ and $k$.
Since the incident angle should be smaller than $90^\circ$,
the wavenumber in the $x$ direction is limited as
\begin{align}
  k \leq \sqrt{\epsilon^p} k_0.
\end{align}
Also, note that any excitations cannot go beyond the light line ($\omega \leq ck$).
Therefore, the induced magnetization which can be excited in the Kretschmann configuration is only between the two diagonal dashed lines in \figref{fig:magnetization}.
There is a peak in the region, which is contribution from the surface plasmon mode at the air-metal interface.
If we replot the induced magnetization as a function of the incident angle for a given incident frequency (\figref{fig:magnetization_angle}),
we can clearly see that there is a peak at the SP angle for a frequency of $\omega=2\pi c/(655\times10^{-9})=2.88\times10^{15}\ \mathrm{s^{-1}}$,
\begin{align*}
\theta_\mathrm{sp}=\arcsin\sqrt{\cfrac{\epsilon^m}{\epsilon^a+\epsilon^m}} \approx 43^\circ.
\end{align*}

\section{Diffusive electron spin dynamics in the metal}
In this section,
we analyze the electron spin dynamics in the inhomogeneous magnetic field by using the spin diffusion equation,
which can be derived from the framework of Boltzmann theory and useful to analyze the electron spin transport in metallic systems \cite{johnson1985interfacial,johnson1988spin,takahashi2008spin}.
As discussed in the literature \cite{matsuo2013mechanical, matsuo2017theory}, we need a source term in the diffusion equation to take two kinds of processes, spontaneous diffusion and induced diffusion, into consideration,
\begin{align}
  \left(\frac{\partial}{\partial t} - D_s \nabla^2 + \frac{1}{\tau}\right)\delta \mu = \rho_0 e D_s \nabla \cdot \bm{j}_s^\mathrm{sou},
\end{align}
where spin accumulation $\delta \mu$ is a potential for the electron spin, 
and its gradient drives electron spin currents, $\bm{j}_s \propto \nabla \delta \mu$.
Our source term is given by the gradient of the inhomogeneous magnetization,
\begin{align}
  \bm{j}_s^\mathrm{sou} &= \frac{\hbar \mu_0}{m \rho_0} \nabla M_y.
  \label{eq:j_s^sou}
\end{align}
Here, $D_s = {\lambda_s}^2/\tau$ is the diffusion constant,
$\lambda_s$ and $\tau$ are the spin diffusion length and the spin relaxation time in the metal,
and $\rho_0$ is the resistivity of the metal.
Eq. \eqref{eq:j_s^sou} is a spin current driven by the Stern-Gerlach-type mechanism as it is proportional to the gradient of effective magnetic field.
In the original experiment by Gerlach and Stern,
charge neutral beam was utilized in order to eliminate the Lorentz force effect \cite{gerlach1922experimentelle,mott1929scattering}.
Likewise, there is no charge transport in our proposed system.

\par
At the steady state ($\partial/\partial t = 0$),
we obtain the space evolution equation of the spin accumulation,
\begin{align}
  \nabla^2 \delta \mu &= \frac{1}{{\lambda_s}^2} \delta \mu + \frac{\hbar e}{m} \mu_0 \nabla^2 M_y.
\end{align}
There are two characteristic lengths in this equation.
One is the spin diffusion length $\lambda_s$, 
which characterises the spontaneous diffusion process and is typically in the order of $10\ \mathrm{nm}$ in the metallic systems.
The other is the penetration length of the plasmonic mode $\mathfrak{Im}(K^m)$, 
which is hidden in the gradient operator in front of the magnetization $M_y$, 
and characterises the diffusion process induced optically.

\par
The particular solution of this inhomogeneous differential equation is
\begin{align}
  {\delta \mu}^\mathrm{sp} &= \sum_\sigma 
  A_\sigma e^{-2\sigma\mathfrak{Im}(K^m)z},
  \label{eq:particular}
  \\
  A_\sigma &= \sigma \frac{2k\mathfrak{Im}(K^m)}{\omega^2\epsilon^m/c^2}
  L(k,\omega,\lambda_s) \gamma \mu_B 
  \mu_0 \frac{d\epsilon^m}{d\omega} u_\sigma(k,\omega),
  \label{eq:A_sigma}
\end{align}
where $\gamma = \frac{e}{2m}$ is the gyromagnetic ratio,
$\mu_B = \frac{e\hbar}{2m}$ is the Bohr magneton,
and we have defined
\begin{align}
  L(k,\omega,\lambda_s) &= \frac{\{2\mathfrak{Im}(K^m)\lambda_s\}^2}{\{2\mathfrak{Im}(K^m) \lambda_s \}^2 - 1},
  \label{eq:Lorentz_factor}
  \\
  u_\sigma(k,\omega) &= \epsilon_0 |E_{p\sigma}^m|^2.
\end{align}
On the other hand, the general solution of the corresponding homogeneous equation is
\begin{align}
  {\delta \mu}^0 &= \sum_\sigma B_\sigma e^{-\sigma z/\lambda_s}.
  \label{eq:general}
\end{align}

\begin{figure*}[htbp]
  \subfloat[$j_s(\omega,\theta_\mathrm{in},\lambda_s = 10\ \mathrm{nm})$\label{subfig:js_omega_theta}]{%
    \includegraphics[width=0.33\linewidth]{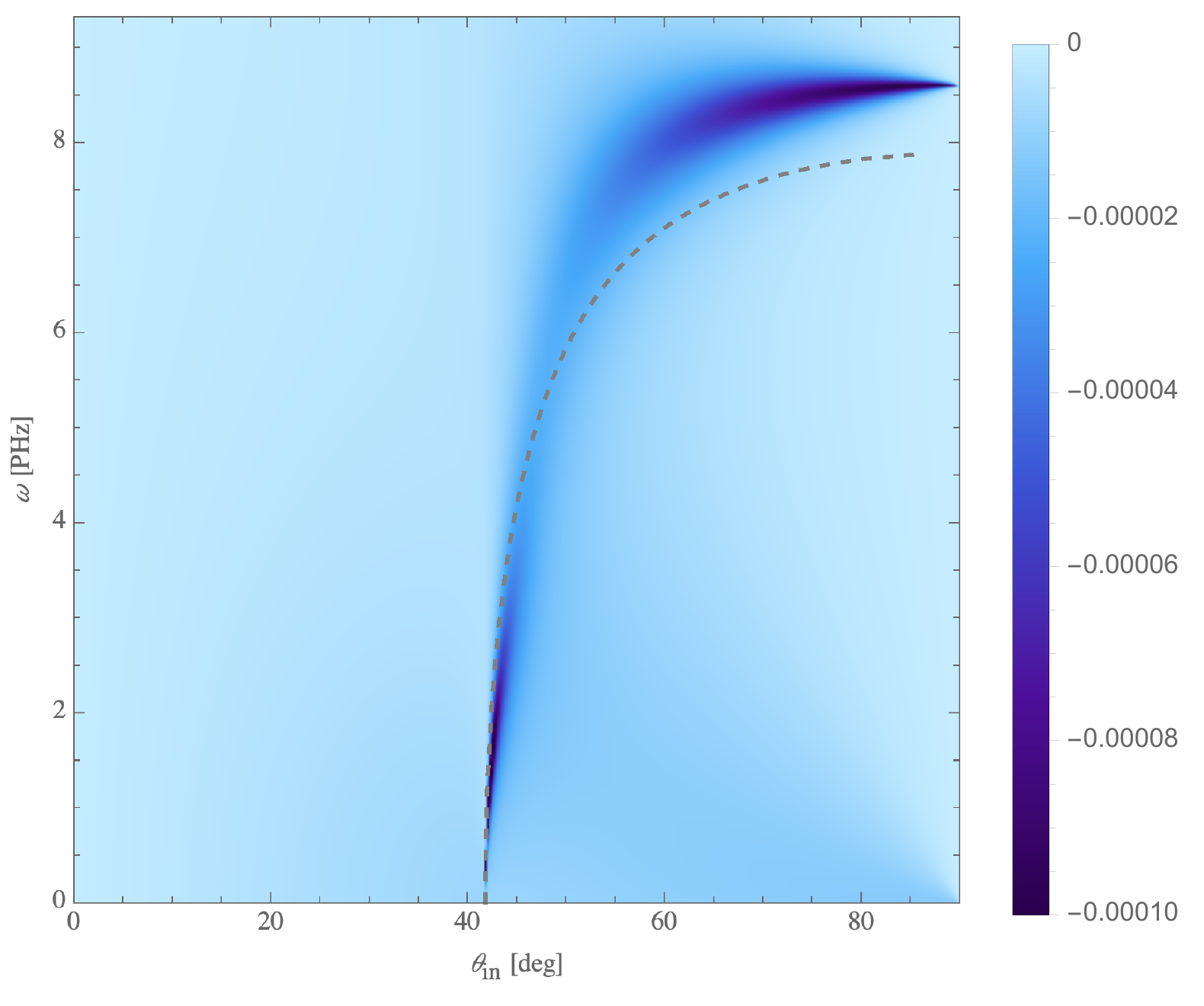}%
}\hfill
\subfloat[$j_s(\omega,\theta_\mathrm{in}=\theta_\mathrm{sp},\lambda_s)$\label{subfig:js_lambdas_omega}]{%
    \includegraphics[width=0.33\linewidth]{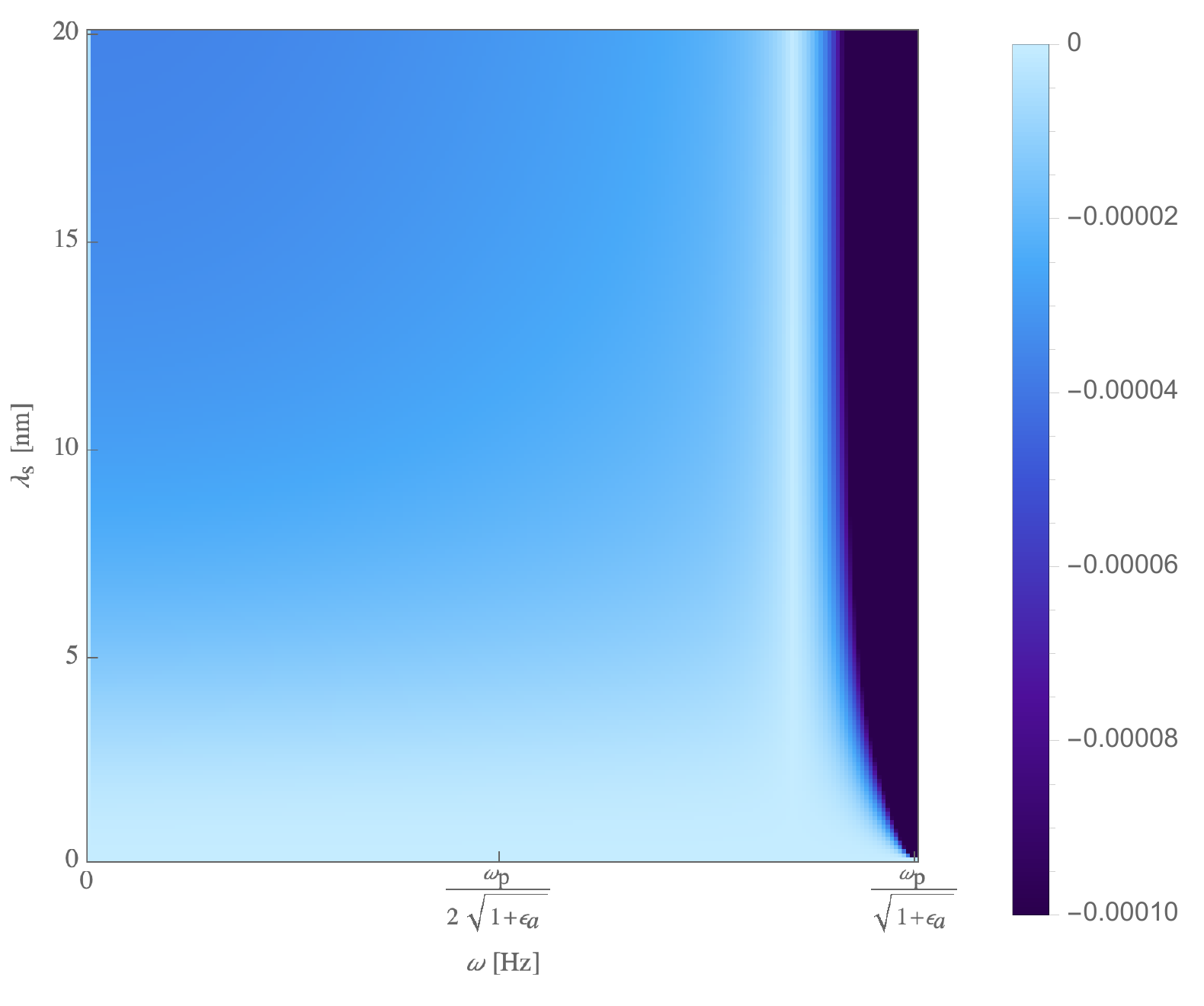}%
}\hfill
\subfloat[$j_s(\omega=\omega_\mathrm{p}/2\sqrt{1+\epsilon^a},\theta_\mathrm{in},\lambda_s)$\label{subfig:js_theta_lambdas}]{%
    \includegraphics[width=0.33\linewidth]{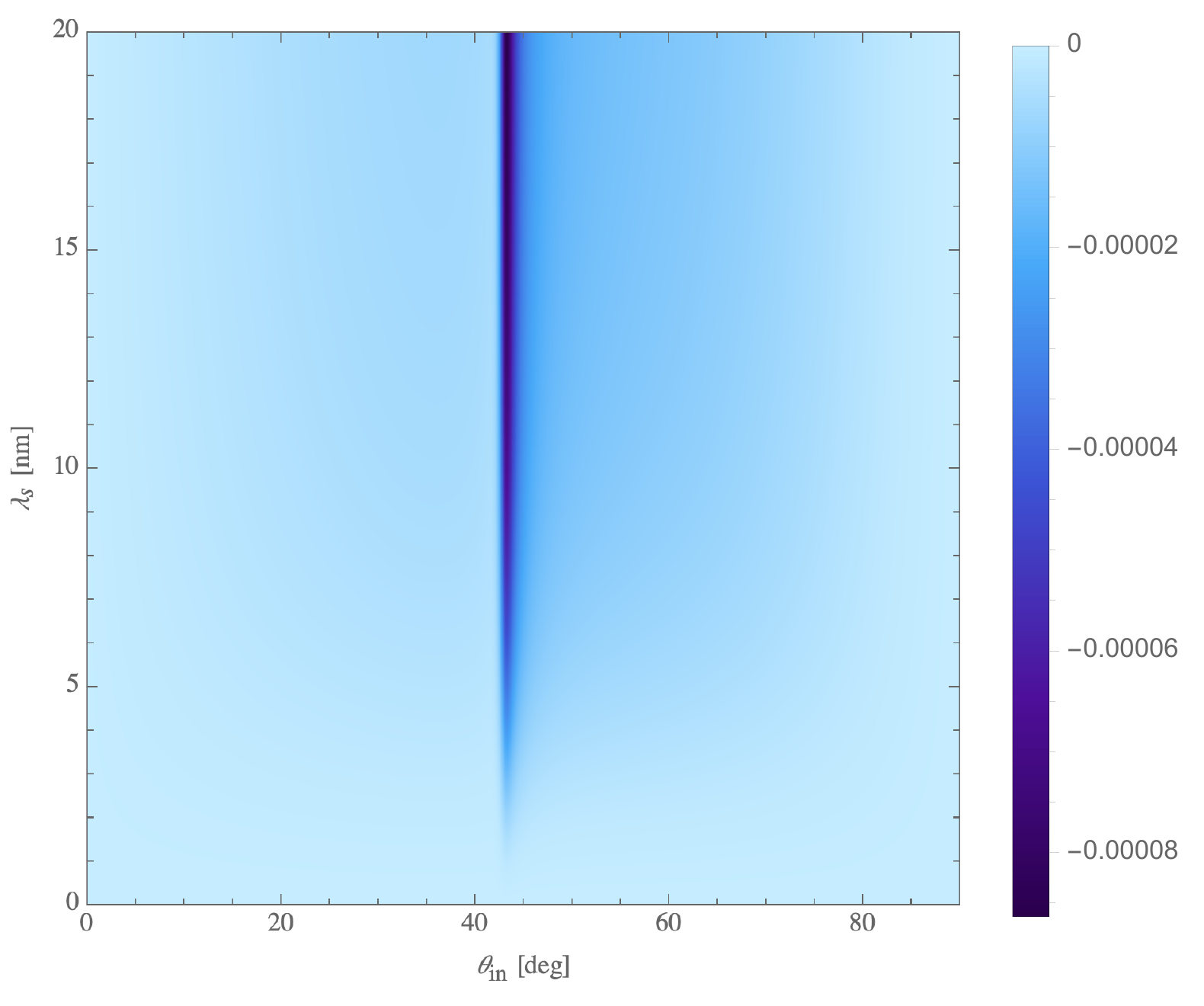}%
}    
  \caption{
    Optically generated spin current $j_s$ at $z=0$ in the Kretschmann configuration.
    (a) colormap of $j_s$ as a function of the incident frequency $\omega$ and the incident angle $\theta_\mathrm{in}$ for a fixed spin diffusion length $\lambda_s$.
    The grey dashed line indicates the SP angle $\theta_\mathrm{sp}(\omega)$.
    We can see that the peak of the spin current density is around the dashed curve,
    which implies the spin current is driven by the SP.
    The peak slightly deviate from the SP angle condition around the right top region, 
    where the spontaneous diffusion is comparable to the induced diffusion.
    (b) colormap of $j_s$ as a function of the incident frequency $\omega$ and the spin diffusion length $\lambda_s$ at the SP angle of incidence $\theta_\mathrm{in} = \theta_\mathrm{sp}$.
    (c) colormap of $j_s$ as a function of the incident angle $\theta_\mathrm{in}$ and the spin diffusion length at a given frequency.
    To generate these plots, we set $\rho_0 = 12.8 \times 10^{-8}$ (the resistivity of gold),
    and other parameters are the same as the previous figures.
    Note that we set the lower bounds of the plot ranges $-0.1\ \mathrm{mA/m^2}$ so that the peaks can be seen clearly.
  }
  \label{fig:js}
\end{figure*}

\par
The coefficients, $B_\sigma$, are determined by the boundary condition that the derivative of the total spin accumulation (i.e., the diffusive spin current) vanishes at the two boundaries,
\begin{align}
 0 = \left.\nabla \delta \mu \right|_{z=\pm d/2} = \left.\nabla ({\delta \mu}^0 + {\delta \mu}^\mathrm{sp})\right|_{z=\pm d/2}.
\end{align}
This yields simultaneous equations,
\begin{align*}
  &\begin{pmatrix}
    e^{-\frac{d}{2\lambda_s}} & -e^{+\frac{d}{2\lambda_s}}\\
    e^{+\frac{d}{2\lambda_s}} & -e^{-\frac{d}{2\lambda_s}}
  \end{pmatrix}
  \begin{pmatrix}
    B_+\\
    B_-
  \end{pmatrix}
  \notag \\
  &=
  2\mathfrak{Im}(K^m)\lambda_s
  \begin{pmatrix}
    -e^{-\mathfrak{Im}(K^m)d} & e^{+\mathfrak{Im}(K^m)d}\\
    -e^{+\mathfrak{Im}(K^m)d} & e^{-\mathfrak{Im}(K^m)d}
  \end{pmatrix}
  \begin{pmatrix}
    A_+\\
    A_-
  \end{pmatrix},
\end{align*}
and then, we get
\begin{align}
  \begin{pmatrix}
    B_+\\
    B_-
  \end{pmatrix}
  &= -\frac{\lambda_s}{d}\frac{X + Y}{\sinh (X - Y)}
  \begin{pmatrix}
    \sinh X & \sinh Y\\
    \sinh Y & \sinh X\\
  \end{pmatrix}
  \begin{pmatrix}
    A_+\\
    A_-
  \end{pmatrix},
\end{align}
where we have defined
\begin{align*}
  X &= \frac{d}{2\lambda_s}\{2\mathfrak{Im}(K^m)\lambda_s + 1\},\\
  Y &= \frac{d}{2\lambda_s}\{2\mathfrak{Im}(K^m)\lambda_s - 1\}.
\end{align*}

\par
Finally, we get the spin accumulation,
\begin{align}
  \delta \mu &= \sum_\sigma A_\sigma e^{-2\sigma \mathfrak{Im}(K^m) z} + B_\sigma e^{-\sigma z/\lambda_s},
\end{align}
and the spin current driven by the spin accumulation,
\begin{align}
  \bm{j}_s 
  &= \frac{1}{\rho_0 e} \nabla \delta \mu 
  \\          
  &= -\frac{1}{\rho_0 e}\sum_\sigma \sigma
  \left(
    2\mathfrak{Im}(K^m) A_\sigma e^{-2\sigma\mathfrak{Im}(K^m)z}
    \right.
    \notag \\ 
  &\hspace{7em}
    \left.
    +\frac{1}{\lambda_s} B_\sigma e^{-\sigma z/\lambda_s}
  \right)\bm{u}_z.
\end{align}
In \figref{fig:js},
the spin current is shown as a function of the incident frequency $\omega$, the incident angle $\theta_\mathrm{in}$, and the spin diffusion length of the metal $\lambda_s$.
From \figref{subfig:js_omega_theta}, 
we can see that the spin current peaks around the SP angle of incidence.
The peak deviates from the SP angle condition at high frequency region.
This is because the decay length of the SP $\mathfrak{Im}(K^m)$ and the spin diffusion length $\lambda_s$ are comparable in that region,
and the spontaneous diffusion become comparable with the induced diffusion.

\section{Conclusion}
To sum up,
in this work,
we have calculated a plasmonic system coupled to external laser drive where electron spin currents are generated.

\par
In order to perform the analysis of electromagnetic field in the system,
the scattering matrix formalism is utilized.
Using the data provided by the field analysis, we calculated magnetization induced by plasma oscillation in the metal.
Since the magnetization inherits the frequency dependence, the incident angle dependence, and the spatial distribution from the plasmonic modes, 
it acts as inhomogeneous effective magnetic field which peaks at a condition where the surface plasmon mode is excited by the external drive and causes the Stern-Gerlach effect to generate electron spin current.

\par
We have also solved spin diffusion equation to analyze the spin current in detail.
The optically induced spin current can be detected by the inverse spin Hall measurement,
where electrodes attached at the both ends of the metal film detect inverse spin Hall voltage stemming from the spin current in the thickness direction (see, for example, \cite{ando2011inverse} for the measurement scheme).
By modulating the polarization of the external drive between $s$ and $p$ polarisation states as in the literature \cite{lindemann2019ultrafast},
we could perform the lock-in-type measurement to discriminate optically induced spin currents from other parasitic effects such as local heating.
Note that while we have focused on and analyzed the diffusive spin transport, 
there will be ballistic transport and transient regime.
In order to consider these kinds of transport,
we need to go back to the framework of Boltzmann theory as they did in the literature \cite{fabian2002spin}.
We leave it for future works.

\par
Our theoretical studies here are feasible for experimental demonstration and commit to further research on the interface between optics and spintronics.

\begin{acknowledgments}
  We deeply thank Dr. Junji Fujimoto for a very helpful discussion.
  We also thank Prof. Yukio Nozaki and members in his research group for fruitful discussions.
  MM is partially Supported by the Priority Program of Chinese Academy of Sciences, Grant No. XDB28000000.
\end{acknowledgments}


\begin{thebibliography}{37}%
\makeatletter
\providecommand \@ifxundefined [1]{%
 \@ifx{#1\undefined}
}%
\providecommand \@ifnum [1]{%
 \ifnum #1\expandafter \@firstoftwo
 \else \expandafter \@secondoftwo
 \fi
}%
\providecommand \@ifx [1]{%
 \ifx #1\expandafter \@firstoftwo
 \else \expandafter \@secondoftwo
 \fi
}%
\providecommand \natexlab [1]{#1}%
\providecommand \enquote  [1]{``#1''}%
\providecommand \bibnamefont  [1]{#1}%
\providecommand \bibfnamefont [1]{#1}%
\providecommand \citenamefont [1]{#1}%
\providecommand \href@noop [0]{\@secondoftwo}%
\providecommand \href [0]{\begingroup \@sanitize@url \@href}%
\providecommand \@href[1]{\@@startlink{#1}\@@href}%
\providecommand \@@href[1]{\endgroup#1\@@endlink}%
\providecommand \@sanitize@url [0]{\catcode `\\12\catcode `\$12\catcode
  `\&12\catcode `\#12\catcode `\^12\catcode `\_12\catcode `\%12\relax}%
\providecommand \@@startlink[1]{}%
\providecommand \@@endlink[0]{}%
\providecommand \url  [0]{\begingroup\@sanitize@url \@url }%
\providecommand \@url [1]{\endgroup\@href {#1}{\urlprefix }}%
\providecommand \urlprefix  [0]{URL }%
\providecommand \Eprint [0]{\href }%
\providecommand \doibase [0]{https://doi.org/}%
\providecommand \selectlanguage [0]{\@gobble}%
\providecommand \bibinfo  [0]{\@secondoftwo}%
\providecommand \bibfield  [0]{\@secondoftwo}%
\providecommand \translation [1]{[#1]}%
\providecommand \BibitemOpen [0]{}%
\providecommand \bibitemStop [0]{}%
\providecommand \bibitemNoStop [0]{.\EOS\space}%
\providecommand \EOS [0]{\spacefactor3000\relax}%
\providecommand \BibitemShut  [1]{\csname bibitem#1\endcsname}%
\let\auto@bib@innerbib\@empty
\bibitem [{\citenamefont {Flipse}\ \emph {et~al.}(2014)\citenamefont {Flipse},
  \citenamefont {Dejene}, \citenamefont {Wagenaar}, \citenamefont {Bauer},
  \citenamefont {Youssef},\ and\ \citenamefont
  {Van~Wees}}]{flipse2014observation}%
  \BibitemOpen
  \bibfield  {author} {\bibinfo {author} {\bibfnamefont {J.}~\bibnamefont
  {Flipse}}, \bibinfo {author} {\bibfnamefont {F.}~\bibnamefont {Dejene}},
  \bibinfo {author} {\bibfnamefont {D.}~\bibnamefont {Wagenaar}}, \bibinfo
  {author} {\bibfnamefont {G.}~\bibnamefont {Bauer}}, \bibinfo {author}
  {\bibfnamefont {J.~B.}\ \bibnamefont {Youssef}},\ and\ \bibinfo {author}
  {\bibfnamefont {B.}~\bibnamefont {Van~Wees}},\ }\bibfield  {title} {\bibinfo
  {title} {Observation of the spin {Peltier} effect for magnetic insulators},\
  }\href@noop {} {\bibfield  {journal} {\bibinfo  {journal} {Phys. Rev. Lett.}\
  }\textbf {\bibinfo {volume} {113}},\ \bibinfo {pages} {027601} (\bibinfo
  {year} {2014})}\BibitemShut {NoStop}%
\bibitem [{\citenamefont {Daimon}\ \emph {et~al.}(2016)\citenamefont {Daimon},
  \citenamefont {Iguchi}, \citenamefont {Hioki}, \citenamefont {Saitoh},\ and\
  \citenamefont {Uchida}}]{daimon2016thermal}%
  \BibitemOpen
  \bibfield  {author} {\bibinfo {author} {\bibfnamefont {S.}~\bibnamefont
  {Daimon}}, \bibinfo {author} {\bibfnamefont {R.}~\bibnamefont {Iguchi}},
  \bibinfo {author} {\bibfnamefont {T.}~\bibnamefont {Hioki}}, \bibinfo
  {author} {\bibfnamefont {E.}~\bibnamefont {Saitoh}},\ and\ \bibinfo {author}
  {\bibfnamefont {K.-i.}\ \bibnamefont {Uchida}},\ }\bibfield  {title}
  {\bibinfo {title} {Thermal imaging of spin {Peltier} effect},\ }\href@noop {}
  {\bibfield  {journal} {\bibinfo  {journal} {Nat. Commun.}\ }\textbf {\bibinfo
  {volume} {7}},\ \bibinfo {pages} {1} (\bibinfo {year} {2016})}\BibitemShut
  {NoStop}%
\bibitem [{\citenamefont {Zolfagharkhani}\ \emph {et~al.}(2008)\citenamefont
  {Zolfagharkhani}, \citenamefont {Gaidarzhy}, \citenamefont {Degiovanni},
  \citenamefont {Kettemann}, \citenamefont {Fulde},\ and\ \citenamefont
  {Mohanty}}]{zolfagharkhani2008nanomechanical}%
  \BibitemOpen
  \bibfield  {author} {\bibinfo {author} {\bibfnamefont {G.}~\bibnamefont
  {Zolfagharkhani}}, \bibinfo {author} {\bibfnamefont {A.}~\bibnamefont
  {Gaidarzhy}}, \bibinfo {author} {\bibfnamefont {P.}~\bibnamefont
  {Degiovanni}}, \bibinfo {author} {\bibfnamefont {S.}~\bibnamefont
  {Kettemann}}, \bibinfo {author} {\bibfnamefont {P.}~\bibnamefont {Fulde}},\
  and\ \bibinfo {author} {\bibfnamefont {P.}~\bibnamefont {Mohanty}},\
  }\bibfield  {title} {\bibinfo {title} {Nanomechanical detection of itinerant
  electron spin flip},\ }\href@noop {} {\bibfield  {journal} {\bibinfo
  {journal} {Nat. Nanotechnol.}\ }\textbf {\bibinfo {volume} {3}},\ \bibinfo
  {pages} {720} (\bibinfo {year} {2008})}\BibitemShut {NoStop}%
\bibitem [{\citenamefont {Harii}\ \emph {et~al.}(2019)\citenamefont {Harii},
  \citenamefont {Seo}, \citenamefont {Tsutsumi}, \citenamefont {Chudo},
  \citenamefont {Oyanagi}, \citenamefont {Matsuo}, \citenamefont {Shiomi},
  \citenamefont {Ono}, \citenamefont {Maekawa},\ and\ \citenamefont
  {Saitoh}}]{harii2019spin}%
  \BibitemOpen
  \bibfield  {author} {\bibinfo {author} {\bibfnamefont {K.}~\bibnamefont
  {Harii}}, \bibinfo {author} {\bibfnamefont {Y.-J.}\ \bibnamefont {Seo}},
  \bibinfo {author} {\bibfnamefont {Y.}~\bibnamefont {Tsutsumi}}, \bibinfo
  {author} {\bibfnamefont {H.}~\bibnamefont {Chudo}}, \bibinfo {author}
  {\bibfnamefont {K.}~\bibnamefont {Oyanagi}}, \bibinfo {author} {\bibfnamefont
  {M.}~\bibnamefont {Matsuo}}, \bibinfo {author} {\bibfnamefont
  {Y.}~\bibnamefont {Shiomi}}, \bibinfo {author} {\bibfnamefont
  {T.}~\bibnamefont {Ono}}, \bibinfo {author} {\bibfnamefont {S.}~\bibnamefont
  {Maekawa}},\ and\ \bibinfo {author} {\bibfnamefont {E.}~\bibnamefont
  {Saitoh}},\ }\bibfield  {title} {\bibinfo {title} {Spin {Seebeck} mechanical
  force},\ }\href@noop {} {\bibfield  {journal} {\bibinfo  {journal} {Nat.
  Commun.}\ }\textbf {\bibinfo {volume} {10}},\ \bibinfo {pages} {1} (\bibinfo
  {year} {2019})}\BibitemShut {NoStop}%
\bibitem [{\citenamefont {Vrijen}\ \emph {et~al.}(2000)\citenamefont {Vrijen},
  \citenamefont {Yablonovitch}, \citenamefont {Wang}, \citenamefont {Jiang},
  \citenamefont {Balandin}, \citenamefont {Roychowdhury}, \citenamefont {Mor},\
  and\ \citenamefont {DiVincenzo}}]{vrijen2000electron}%
  \BibitemOpen
  \bibfield  {author} {\bibinfo {author} {\bibfnamefont {R.}~\bibnamefont
  {Vrijen}}, \bibinfo {author} {\bibfnamefont {E.}~\bibnamefont
  {Yablonovitch}}, \bibinfo {author} {\bibfnamefont {K.}~\bibnamefont {Wang}},
  \bibinfo {author} {\bibfnamefont {H.~W.}\ \bibnamefont {Jiang}}, \bibinfo
  {author} {\bibfnamefont {A.}~\bibnamefont {Balandin}}, \bibinfo {author}
  {\bibfnamefont {V.}~\bibnamefont {Roychowdhury}}, \bibinfo {author}
  {\bibfnamefont {T.}~\bibnamefont {Mor}},\ and\ \bibinfo {author}
  {\bibfnamefont {D.}~\bibnamefont {DiVincenzo}},\ }\bibfield  {title}
  {\bibinfo {title} {Electron-spin-resonance transistors for quantum computing
  in silicon-germanium heterostructures},\ }\href@noop {} {\bibfield  {journal}
  {\bibinfo  {journal} {Phys. Rev. A}\ }\textbf {\bibinfo {volume} {62}},\
  \bibinfo {pages} {012306} (\bibinfo {year} {2000})}\BibitemShut {NoStop}%
\bibitem [{\citenamefont {Wesenberg}\ \emph {et~al.}(2009)\citenamefont
  {Wesenberg}, \citenamefont {Ardavan}, \citenamefont {Briggs}, \citenamefont
  {Morton}, \citenamefont {Schoelkopf}, \citenamefont {Schuster},\ and\
  \citenamefont {M{\o}lmer}}]{wesenberg2009quantum}%
  \BibitemOpen
  \bibfield  {author} {\bibinfo {author} {\bibfnamefont {J.~H.}\ \bibnamefont
  {Wesenberg}}, \bibinfo {author} {\bibfnamefont {A.}~\bibnamefont {Ardavan}},
  \bibinfo {author} {\bibfnamefont {G.~A.~D.}\ \bibnamefont {Briggs}}, \bibinfo
  {author} {\bibfnamefont {J.~J.}\ \bibnamefont {Morton}}, \bibinfo {author}
  {\bibfnamefont {R.~J.}\ \bibnamefont {Schoelkopf}}, \bibinfo {author}
  {\bibfnamefont {D.~I.}\ \bibnamefont {Schuster}},\ and\ \bibinfo {author}
  {\bibfnamefont {K.}~\bibnamefont {M{\o}lmer}},\ }\bibfield  {title} {\bibinfo
  {title} {Quantum computing with an electron spin ensemble},\ }\href@noop {}
  {\bibfield  {journal} {\bibinfo  {journal} {Phys. Rev. Lett.}\ }\textbf
  {\bibinfo {volume} {103}},\ \bibinfo {pages} {070502} (\bibinfo {year}
  {2009})}\BibitemShut {NoStop}%
\bibitem [{\citenamefont {Fabian}\ and\ \citenamefont {{Das
  Sarma}}(2002)}]{fabian2002spin}%
  \BibitemOpen
  \bibfield  {author} {\bibinfo {author} {\bibfnamefont {J.}~\bibnamefont
  {Fabian}}\ and\ \bibinfo {author} {\bibfnamefont {S.}~\bibnamefont {{Das
  Sarma}}},\ }\bibfield  {title} {\bibinfo {title} {Spin transport in
  inhomogeneous magnetic fields: A proposal for {Stern}-{Gerlach}-like
  experiments with conduction electrons},\ }\href@noop {} {\bibfield  {journal}
  {\bibinfo  {journal} {Phys. Rev. B}\ }\textbf {\bibinfo {volume} {66}},\
  \bibinfo {pages} {024436} (\bibinfo {year} {2002})}\BibitemShut {NoStop}%
\bibitem [{\citenamefont {{\v{Z}}uti{\'c}}\ \emph {et~al.}(2002)\citenamefont
  {{\v{Z}}uti{\'c}}, \citenamefont {Fabian},\ and\ \citenamefont {{Das
  Sarma}}}]{vzutic2002spin}%
  \BibitemOpen
  \bibfield  {author} {\bibinfo {author} {\bibfnamefont {I.}~\bibnamefont
  {{\v{Z}}uti{\'c}}}, \bibinfo {author} {\bibfnamefont {J.}~\bibnamefont
  {Fabian}},\ and\ \bibinfo {author} {\bibfnamefont {S.}~\bibnamefont {{Das
  Sarma}}},\ }\bibfield  {title} {\bibinfo {title} {Spin-polarized transport in
  inhomogeneous magnetic semiconductors: theory of magnetic/nonmagnetic p- n
  junctions},\ }\href@noop {} {\bibfield  {journal} {\bibinfo  {journal} {Phys.
  Rev. Lett.}\ }\textbf {\bibinfo {volume} {88}},\ \bibinfo {pages} {066603}
  (\bibinfo {year} {2002})}\BibitemShut {NoStop}%
\bibitem [{\citenamefont {Fabian}\ \emph {et~al.}(2002)\citenamefont {Fabian},
  \citenamefont {{\v{Z}}uti{\'c}},\ and\ \citenamefont {{Das
  Sarma}}}]{fabian2002theory}%
  \BibitemOpen
  \bibfield  {author} {\bibinfo {author} {\bibfnamefont {J.}~\bibnamefont
  {Fabian}}, \bibinfo {author} {\bibfnamefont {I.}~\bibnamefont
  {{\v{Z}}uti{\'c}}},\ and\ \bibinfo {author} {\bibfnamefont {S.}~\bibnamefont
  {{Das Sarma}}},\ }\bibfield  {title} {\bibinfo {title} {Theory of
  spin-polarized bipolar transport in magnetic p- n junctions},\ }\href@noop {}
  {\bibfield  {journal} {\bibinfo  {journal} {Phys. Rev. B}\ }\textbf {\bibinfo
  {volume} {66}},\ \bibinfo {pages} {165301} (\bibinfo {year}
  {2002})}\BibitemShut {NoStop}%
\bibitem [{\citenamefont {Takahashi}\ \emph {et~al.}(2016)\citenamefont
  {Takahashi}, \citenamefont {Matsuo}, \citenamefont {Ono}, \citenamefont
  {Harii}, \citenamefont {Chudo}, \citenamefont {Okayasu}, \citenamefont
  {Ieda}, \citenamefont {Takahashi}, \citenamefont {Maekawa},\ and\
  \citenamefont {Saitoh}}]{takahashi2016spin}%
  \BibitemOpen
  \bibfield  {author} {\bibinfo {author} {\bibfnamefont {R.}~\bibnamefont
  {Takahashi}}, \bibinfo {author} {\bibfnamefont {M.}~\bibnamefont {Matsuo}},
  \bibinfo {author} {\bibfnamefont {M.}~\bibnamefont {Ono}}, \bibinfo {author}
  {\bibfnamefont {K.}~\bibnamefont {Harii}}, \bibinfo {author} {\bibfnamefont
  {H.}~\bibnamefont {Chudo}}, \bibinfo {author} {\bibfnamefont
  {S.}~\bibnamefont {Okayasu}}, \bibinfo {author} {\bibfnamefont
  {J.}~\bibnamefont {Ieda}}, \bibinfo {author} {\bibfnamefont {S.}~\bibnamefont
  {Takahashi}}, \bibinfo {author} {\bibfnamefont {S.}~\bibnamefont {Maekawa}},\
  and\ \bibinfo {author} {\bibfnamefont {E.}~\bibnamefont {Saitoh}},\
  }\bibfield  {title} {\bibinfo {title} {Spin hydrodynamic generation},\
  }\href@noop {} {\bibfield  {journal} {\bibinfo  {journal} {Nat. Phys.}\
  }\textbf {\bibinfo {volume} {12}},\ \bibinfo {pages} {52} (\bibinfo {year}
  {2016})}\BibitemShut {NoStop}%
\bibitem [{\citenamefont {Kobayashi}\ \emph {et~al.}(2017)\citenamefont
  {Kobayashi}, \citenamefont {Yoshikawa}, \citenamefont {Matsuo}, \citenamefont
  {Iguchi}, \citenamefont {Maekawa}, \citenamefont {Saitoh},\ and\
  \citenamefont {Nozaki}}]{kobayashi2017spin}%
  \BibitemOpen
  \bibfield  {author} {\bibinfo {author} {\bibfnamefont {D.}~\bibnamefont
  {Kobayashi}}, \bibinfo {author} {\bibfnamefont {T.}~\bibnamefont
  {Yoshikawa}}, \bibinfo {author} {\bibfnamefont {M.}~\bibnamefont {Matsuo}},
  \bibinfo {author} {\bibfnamefont {R.}~\bibnamefont {Iguchi}}, \bibinfo
  {author} {\bibfnamefont {S.}~\bibnamefont {Maekawa}}, \bibinfo {author}
  {\bibfnamefont {E.}~\bibnamefont {Saitoh}},\ and\ \bibinfo {author}
  {\bibfnamefont {Y.}~\bibnamefont {Nozaki}},\ }\bibfield  {title} {\bibinfo
  {title} {Spin current generation using a surface acoustic wave generated via
  spin-rotation coupling},\ }\href@noop {} {\bibfield  {journal} {\bibinfo
  {journal} {Phys. Rev. Lett.}\ }\textbf {\bibinfo {volume} {119}},\ \bibinfo
  {pages} {077202} (\bibinfo {year} {2017})}\BibitemShut {NoStop}%
\bibitem [{\citenamefont {Okano}\ \emph {et~al.}(2019)\citenamefont {Okano},
  \citenamefont {Matsuo}, \citenamefont {Ohnuma}, \citenamefont {Maekawa},\
  and\ \citenamefont {Nozaki}}]{okano2019nonreciprocal}%
  \BibitemOpen
  \bibfield  {author} {\bibinfo {author} {\bibfnamefont {G.}~\bibnamefont
  {Okano}}, \bibinfo {author} {\bibfnamefont {M.}~\bibnamefont {Matsuo}},
  \bibinfo {author} {\bibfnamefont {Y.}~\bibnamefont {Ohnuma}}, \bibinfo
  {author} {\bibfnamefont {S.}~\bibnamefont {Maekawa}},\ and\ \bibinfo {author}
  {\bibfnamefont {Y.}~\bibnamefont {Nozaki}},\ }\bibfield  {title} {\bibinfo
  {title} {Nonreciprocal spin current generation in surface-oxidized copper
  films},\ }\href@noop {} {\bibfield  {journal} {\bibinfo  {journal} {Phys.
  Rev. Lett.}\ }\textbf {\bibinfo {volume} {122}},\ \bibinfo {pages} {217701}
  (\bibinfo {year} {2019})}\BibitemShut {NoStop}%
\bibitem [{\citenamefont {Maier}(2007)}]{maier2007plasmonics}%
  \BibitemOpen
  \bibfield  {author} {\bibinfo {author} {\bibfnamefont {S.~A.}\ \bibnamefont
  {Maier}},\ }\href@noop {} {\emph {\bibinfo {title} {Plasmonics: fundamentals
  and applications}}}\ (\bibinfo  {publisher} {Springer Science \& Business
  Media},\ \bibinfo {address} {Berlin},\ \bibinfo {year} {2007})\BibitemShut
  {NoStop}%
\bibitem [{\citenamefont {Uchida}\ \emph {et~al.}(2015)\citenamefont {Uchida},
  \citenamefont {Adachi}, \citenamefont {Kikuchi}, \citenamefont {Ito},
  \citenamefont {Qiu}, \citenamefont {Maekawa},\ and\ \citenamefont
  {Saitoh}}]{uchida2015generation}%
  \BibitemOpen
  \bibfield  {author} {\bibinfo {author} {\bibfnamefont {K.}~\bibnamefont
  {Uchida}}, \bibinfo {author} {\bibfnamefont {H.}~\bibnamefont {Adachi}},
  \bibinfo {author} {\bibfnamefont {D.}~\bibnamefont {Kikuchi}}, \bibinfo
  {author} {\bibfnamefont {S.}~\bibnamefont {Ito}}, \bibinfo {author}
  {\bibfnamefont {Z.}~\bibnamefont {Qiu}}, \bibinfo {author} {\bibfnamefont
  {S.}~\bibnamefont {Maekawa}},\ and\ \bibinfo {author} {\bibfnamefont
  {E.}~\bibnamefont {Saitoh}},\ }\bibfield  {title} {\bibinfo {title}
  {Generation of spin currents by surface plasmon resonance},\ }\href@noop {}
  {\bibfield  {journal} {\bibinfo  {journal} {Nat. Commun.}\ }\textbf {\bibinfo
  {volume} {6}},\ \bibinfo {pages} {5910} (\bibinfo {year} {2015})}\BibitemShut
  {NoStop}%
\bibitem [{\citenamefont {Ishii}\ \emph {et~al.}(2017)\citenamefont {Ishii},
  \citenamefont {Uchida}, \citenamefont {Dao}, \citenamefont {Wada},
  \citenamefont {Saitoh},\ and\ \citenamefont {Nagao}}]{ishii2017wavelength}%
  \BibitemOpen
  \bibfield  {author} {\bibinfo {author} {\bibfnamefont {S.}~\bibnamefont
  {Ishii}}, \bibinfo {author} {\bibfnamefont {K.-i.}\ \bibnamefont {Uchida}},
  \bibinfo {author} {\bibfnamefont {T.~D.}\ \bibnamefont {Dao}}, \bibinfo
  {author} {\bibfnamefont {Y.}~\bibnamefont {Wada}}, \bibinfo {author}
  {\bibfnamefont {E.}~\bibnamefont {Saitoh}},\ and\ \bibinfo {author}
  {\bibfnamefont {T.}~\bibnamefont {Nagao}},\ }\bibfield  {title} {\bibinfo
  {title} {Wavelength-selective spin-current generator using infrared plasmonic
  metamaterials},\ }\href@noop {} {\bibfield  {journal} {\bibinfo  {journal}
  {APL Photonics}\ }\textbf {\bibinfo {volume} {2}},\ \bibinfo {pages} {106103}
  (\bibinfo {year} {2017})}\BibitemShut {NoStop}%
\bibitem [{\citenamefont {Oue}\ and\ \citenamefont
  {Matsuo}(2020{\natexlab{a}})}]{oue2020electron}%
  \BibitemOpen
  \bibfield  {author} {\bibinfo {author} {\bibfnamefont {D.}~\bibnamefont
  {Oue}}\ and\ \bibinfo {author} {\bibfnamefont {M.}~\bibnamefont {Matsuo}},\
  }\bibfield  {title} {\bibinfo {title} {Electron spin transport driven by
  surface plasmon polaritons},\ }\href@noop {} {\bibfield  {journal} {\bibinfo
  {journal} {Phys. Rev. B}\ }\textbf {\bibinfo {volume} {101}},\ \bibinfo
  {pages} {161404(R)} (\bibinfo {year} {2020}{\natexlab{a}})}\BibitemShut
  {NoStop}%
\bibitem [{\citenamefont {Oue}\ and\ \citenamefont
  {Matsuo}(2020{\natexlab{b}})}]{oue2020effects}%
  \BibitemOpen
  \bibfield  {author} {\bibinfo {author} {\bibfnamefont {D.}~\bibnamefont
  {Oue}}\ and\ \bibinfo {author} {\bibfnamefont {M.}~\bibnamefont {Matsuo}},\
  }\bibfield  {title} {\bibinfo {title} {Effects of surface plasmons on spin
  currents in a thin film system},\ }\href@noop {} {\bibfield  {journal}
  {\bibinfo  {journal} {New J. Phys.}\ }\textbf {\bibinfo {volume} {22}},\
  \bibinfo {pages} {033040} (\bibinfo {year} {2020}{\natexlab{b}})}\BibitemShut
  {NoStop}%
\bibitem [{\citenamefont {Matsuo}\ \emph {et~al.}(2018)\citenamefont {Matsuo},
  \citenamefont {Ohnuma}, \citenamefont {Kato},\ and\ \citenamefont
  {Maekawa}}]{matsuo2018spin}%
  \BibitemOpen
  \bibfield  {author} {\bibinfo {author} {\bibfnamefont {M.}~\bibnamefont
  {Matsuo}}, \bibinfo {author} {\bibfnamefont {Y.}~\bibnamefont {Ohnuma}},
  \bibinfo {author} {\bibfnamefont {T.}~\bibnamefont {Kato}},\ and\ \bibinfo
  {author} {\bibfnamefont {S.}~\bibnamefont {Maekawa}},\ }\bibfield  {title}
  {\bibinfo {title} {Spin current noise of the spin {Seebeck} effect and spin
  pumping},\ }\href@noop {} {\bibfield  {journal} {\bibinfo  {journal} {Phys.
  Rev. Lett.}\ }\textbf {\bibinfo {volume} {120}},\ \bibinfo {pages} {037201}
  (\bibinfo {year} {2018})}\BibitemShut {NoStop}%
\bibitem [{\citenamefont {Kato}\ \emph {et~al.}(2019)\citenamefont {Kato},
  \citenamefont {Ohnuma}, \citenamefont {Matsuo}, \citenamefont {Rech},
  \citenamefont {Jonckheere},\ and\ \citenamefont
  {Martin}}]{kato2019microscopic}%
  \BibitemOpen
  \bibfield  {author} {\bibinfo {author} {\bibfnamefont {T.}~\bibnamefont
  {Kato}}, \bibinfo {author} {\bibfnamefont {Y.}~\bibnamefont {Ohnuma}},
  \bibinfo {author} {\bibfnamefont {M.}~\bibnamefont {Matsuo}}, \bibinfo
  {author} {\bibfnamefont {J.}~\bibnamefont {Rech}}, \bibinfo {author}
  {\bibfnamefont {T.}~\bibnamefont {Jonckheere}},\ and\ \bibinfo {author}
  {\bibfnamefont {T.}~\bibnamefont {Martin}},\ }\bibfield  {title} {\bibinfo
  {title} {Microscopic theory of spin transport at the interface between a
  superconductor and a ferromagnetic insulator},\ }\href@noop {} {\bibfield
  {journal} {\bibinfo  {journal} {Phys. Rev. B}\ }\textbf {\bibinfo {volume}
  {99}},\ \bibinfo {pages} {144411} (\bibinfo {year} {2019})}\BibitemShut
  {NoStop}%
\bibitem [{\citenamefont {Bliokh}\ and\ \citenamefont
  {Nori}(2012)}]{bliokh2012transverse}%
  \BibitemOpen
  \bibfield  {author} {\bibinfo {author} {\bibfnamefont {K.~Y.}\ \bibnamefont
  {Bliokh}}\ and\ \bibinfo {author} {\bibfnamefont {F.}~\bibnamefont {Nori}},\
  }\bibfield  {title} {\bibinfo {title} {Transverse spin of a surface
  polariton},\ }\href@noop {} {\bibfield  {journal} {\bibinfo  {journal} {Phys.
  Rev. A}\ }\textbf {\bibinfo {volume} {85}},\ \bibinfo {pages} {061801}
  (\bibinfo {year} {2012})}\BibitemShut {NoStop}%
\bibitem [{\citenamefont {Bliokh}\ \emph {et~al.}(2014)\citenamefont {Bliokh},
  \citenamefont {Bekshaev},\ and\ \citenamefont
  {Nori}}]{bliokh2014extraordinary}%
  \BibitemOpen
  \bibfield  {author} {\bibinfo {author} {\bibfnamefont {K.~Y.}\ \bibnamefont
  {Bliokh}}, \bibinfo {author} {\bibfnamefont {A.~Y.}\ \bibnamefont
  {Bekshaev}},\ and\ \bibinfo {author} {\bibfnamefont {F.}~\bibnamefont
  {Nori}},\ }\bibfield  {title} {\bibinfo {title} {Extraordinary momentum and
  spin in evanescent waves},\ }\href@noop {} {\bibfield  {journal} {\bibinfo
  {journal} {Nat. Commun.}\ }\textbf {\bibinfo {volume} {5}},\ \bibinfo {pages}
  {3300} (\bibinfo {year} {2014})}\BibitemShut {NoStop}%
\bibitem [{\citenamefont {Bekshaev}\ \emph {et~al.}(2015)\citenamefont
  {Bekshaev}, \citenamefont {Bliokh},\ and\ \citenamefont
  {Nori}}]{bekshaev2015transverse}%
  \BibitemOpen
  \bibfield  {author} {\bibinfo {author} {\bibfnamefont {A.~Y.}\ \bibnamefont
  {Bekshaev}}, \bibinfo {author} {\bibfnamefont {K.~Y.}\ \bibnamefont
  {Bliokh}},\ and\ \bibinfo {author} {\bibfnamefont {F.}~\bibnamefont {Nori}},\
  }\bibfield  {title} {\bibinfo {title} {Transverse spin and momentum in
  two-wave interference},\ }\href@noop {} {\bibfield  {journal} {\bibinfo
  {journal} {Phys. Rev. X}\ }\textbf {\bibinfo {volume} {5}},\ \bibinfo {pages}
  {011039} (\bibinfo {year} {2015})}\BibitemShut {NoStop}%
\bibitem [{\citenamefont {Bliokh}\ \emph {et~al.}(2015)\citenamefont {Bliokh},
  \citenamefont {Smirnova},\ and\ \citenamefont {Nori}}]{bliokh2015quantum}%
  \BibitemOpen
  \bibfield  {author} {\bibinfo {author} {\bibfnamefont {K.~Y.}\ \bibnamefont
  {Bliokh}}, \bibinfo {author} {\bibfnamefont {D.}~\bibnamefont {Smirnova}},\
  and\ \bibinfo {author} {\bibfnamefont {F.}~\bibnamefont {Nori}},\ }\bibfield
  {title} {\bibinfo {title} {Quantum spin {Hall} effect of light},\ }\href@noop
  {} {\bibfield  {journal} {\bibinfo  {journal} {Science}\ }\textbf {\bibinfo
  {volume} {348}},\ \bibinfo {pages} {1448} (\bibinfo {year}
  {2015})}\BibitemShut {NoStop}%
\bibitem [{\citenamefont {Van~Mechelen}\ and\ \citenamefont
  {Jacob}(2016)}]{van2016universal}%
  \BibitemOpen
  \bibfield  {author} {\bibinfo {author} {\bibfnamefont {T.}~\bibnamefont
  {Van~Mechelen}}\ and\ \bibinfo {author} {\bibfnamefont {Z.}~\bibnamefont
  {Jacob}},\ }\bibfield  {title} {\bibinfo {title} {Universal spin-momentum
  locking of evanescent waves},\ }\href@noop {} {\bibfield  {journal} {\bibinfo
   {journal} {Optica}\ }\textbf {\bibinfo {volume} {3}},\ \bibinfo {pages}
  {118} (\bibinfo {year} {2016})}\BibitemShut {NoStop}%
\bibitem [{\citenamefont {Fang}\ and\ \citenamefont
  {Wang}(2017)}]{fang2017intrinsic}%
  \BibitemOpen
  \bibfield  {author} {\bibinfo {author} {\bibfnamefont {L.}~\bibnamefont
  {Fang}}\ and\ \bibinfo {author} {\bibfnamefont {J.}~\bibnamefont {Wang}},\
  }\bibfield  {title} {\bibinfo {title} {Intrinsic transverse spin angular
  momentum of fiber eigenmodes},\ }\href@noop {} {\bibfield  {journal}
  {\bibinfo  {journal} {Phys. Rev. A}\ }\textbf {\bibinfo {volume} {95}},\
  \bibinfo {pages} {053827} (\bibinfo {year} {2017})}\BibitemShut {NoStop}%
\bibitem [{\citenamefont {Oue}(2019)}]{oue2019dissipation}%
  \BibitemOpen
  \bibfield  {author} {\bibinfo {author} {\bibfnamefont {D.}~\bibnamefont
  {Oue}},\ }\bibfield  {title} {\bibinfo {title} {Dissipation effect on optical
  force and torque near interfaces},\ }\href@noop {} {\bibfield  {journal}
  {\bibinfo  {journal} {J. Opt.}\ }\textbf {\bibinfo {volume} {21}},\ \bibinfo
  {pages} {065601} (\bibinfo {year} {2019})}\BibitemShut {NoStop}%
\bibitem [{\citenamefont {Bliokh}\ \emph
  {et~al.}(2017{\natexlab{a}})\citenamefont {Bliokh}, \citenamefont
  {Bekshaev},\ and\ \citenamefont {Nori}}]{bliokh2017optical}%
  \BibitemOpen
  \bibfield  {author} {\bibinfo {author} {\bibfnamefont {K.~Y.}\ \bibnamefont
  {Bliokh}}, \bibinfo {author} {\bibfnamefont {A.~Y.}\ \bibnamefont
  {Bekshaev}},\ and\ \bibinfo {author} {\bibfnamefont {F.}~\bibnamefont
  {Nori}},\ }\bibfield  {title} {\bibinfo {title} {Optical momentum, spin, and
  angular momentum in dispersive media},\ }\href@noop {} {\bibfield  {journal}
  {\bibinfo  {journal} {Phys. Rev. Lett.}\ }\textbf {\bibinfo {volume} {119}},\
  \bibinfo {pages} {073901} (\bibinfo {year} {2017}{\natexlab{a}})}\BibitemShut
  {NoStop}%
\bibitem [{\citenamefont {Bliokh}\ \emph
  {et~al.}(2017{\natexlab{b}})\citenamefont {Bliokh}, \citenamefont
  {Bekshaev},\ and\ \citenamefont {Nori}}]{bliokh2017optical_new_j_phys}%
  \BibitemOpen
  \bibfield  {author} {\bibinfo {author} {\bibfnamefont {K.~Y.}\ \bibnamefont
  {Bliokh}}, \bibinfo {author} {\bibfnamefont {A.~Y.}\ \bibnamefont
  {Bekshaev}},\ and\ \bibinfo {author} {\bibfnamefont {F.}~\bibnamefont
  {Nori}},\ }\bibfield  {title} {\bibinfo {title} {Optical momentum and angular
  momentum in complex media: from the {Abraham}--{Minkowski} debate to unusual
  properties of surface plasmon-polaritons},\ }\href@noop {} {\bibfield
  {journal} {\bibinfo  {journal} {New J. Phys.}\ }\textbf {\bibinfo {volume}
  {19}},\ \bibinfo {pages} {123014} (\bibinfo {year}
  {2017}{\natexlab{b}})}\BibitemShut {NoStop}%
\bibitem [{\citenamefont {Matsuo}\ \emph {et~al.}(2013)\citenamefont {Matsuo},
  \citenamefont {Ieda}, \citenamefont {Harii}, \citenamefont {Saitoh},\ and\
  \citenamefont {Maekawa}}]{matsuo2013mechanical}%
  \BibitemOpen
  \bibfield  {author} {\bibinfo {author} {\bibfnamefont {M.}~\bibnamefont
  {Matsuo}}, \bibinfo {author} {\bibfnamefont {J.}~\bibnamefont {Ieda}},
  \bibinfo {author} {\bibfnamefont {K.}~\bibnamefont {Harii}}, \bibinfo
  {author} {\bibfnamefont {E.}~\bibnamefont {Saitoh}},\ and\ \bibinfo {author}
  {\bibfnamefont {S.}~\bibnamefont {Maekawa}},\ }\bibfield  {title} {\bibinfo
  {title} {Mechanical generation of spin current by spin-rotation coupling},\
  }\href@noop {} {\bibfield  {journal} {\bibinfo  {journal} {Phys. Rev. B}\
  }\textbf {\bibinfo {volume} {87}},\ \bibinfo {pages} {180402} (\bibinfo
  {year} {2013})}\BibitemShut {NoStop}%
\bibitem [{\citenamefont {Matsuo}\ \emph {et~al.}(2017)\citenamefont {Matsuo},
  \citenamefont {Ohnuma},\ and\ \citenamefont {Maekawa}}]{matsuo2017theory}%
  \BibitemOpen
  \bibfield  {author} {\bibinfo {author} {\bibfnamefont {M.}~\bibnamefont
  {Matsuo}}, \bibinfo {author} {\bibfnamefont {Y.}~\bibnamefont {Ohnuma}},\
  and\ \bibinfo {author} {\bibfnamefont {S.}~\bibnamefont {Maekawa}},\
  }\bibfield  {title} {\bibinfo {title} {Theory of spin hydrodynamic
  generation},\ }\href@noop {} {\bibfield  {journal} {\bibinfo  {journal}
  {Phys. Rev. B}\ }\textbf {\bibinfo {volume} {96}},\ \bibinfo {pages} {020401}
  (\bibinfo {year} {2017})}\BibitemShut {NoStop}%
\bibitem [{\citenamefont {Johnson}\ and\ \citenamefont
  {Silsbee}(1985)}]{johnson1985interfacial}%
  \BibitemOpen
  \bibfield  {author} {\bibinfo {author} {\bibfnamefont {M.}~\bibnamefont
  {Johnson}}\ and\ \bibinfo {author} {\bibfnamefont {R.~H.}\ \bibnamefont
  {Silsbee}},\ }\bibfield  {title} {\bibinfo {title} {Interfacial charge-spin
  coupling: Injection and detection of spin magnetization in metals},\
  }\href@noop {} {\bibfield  {journal} {\bibinfo  {journal} {Phys. Rev. Lett.}\
  }\textbf {\bibinfo {volume} {55}},\ \bibinfo {pages} {1790} (\bibinfo {year}
  {1985})}\BibitemShut {NoStop}%
\bibitem [{\citenamefont {Johnson}\ and\ \citenamefont
  {Silsbee}(1988)}]{johnson1988spin}%
  \BibitemOpen
  \bibfield  {author} {\bibinfo {author} {\bibfnamefont {M.}~\bibnamefont
  {Johnson}}\ and\ \bibinfo {author} {\bibfnamefont {R.}~\bibnamefont
  {Silsbee}},\ }\bibfield  {title} {\bibinfo {title} {Spin-injection
  experiment},\ }\href@noop {} {\bibfield  {journal} {\bibinfo  {journal}
  {Phys. Rev. B}\ }\textbf {\bibinfo {volume} {37}},\ \bibinfo {pages} {5326}
  (\bibinfo {year} {1988})}\BibitemShut {NoStop}%
\bibitem [{\citenamefont {Takahashi}\ and\ \citenamefont
  {Maekawa}(2008)}]{takahashi2008spin}%
  \BibitemOpen
  \bibfield  {author} {\bibinfo {author} {\bibfnamefont {S.}~\bibnamefont
  {Takahashi}}\ and\ \bibinfo {author} {\bibfnamefont {S.}~\bibnamefont
  {Maekawa}},\ }\bibfield  {title} {\bibinfo {title} {Spin current, spin
  accumulation and spin {Hall} effect},\ }\href@noop {} {\bibfield  {journal}
  {\bibinfo  {journal} {Sci. Tech. Adv. Mater}\ }\textbf {\bibinfo {volume}
  {9}},\ \bibinfo {pages} {014105} (\bibinfo {year} {2008})}\BibitemShut
  {NoStop}%
\bibitem [{\citenamefont {Gerlach}\ and\ \citenamefont
  {Stern}(1922)}]{gerlach1922experimentelle}%
  \BibitemOpen
  \bibfield  {author} {\bibinfo {author} {\bibfnamefont {W.}~\bibnamefont
  {Gerlach}}\ and\ \bibinfo {author} {\bibfnamefont {O.}~\bibnamefont
  {Stern}},\ }\bibfield  {title} {\bibinfo {title} {Der experimentelle nachweis
  der richtungsquantelung im magnetfeld},\ }\href@noop {} {\bibfield  {journal}
  {\bibinfo  {journal} {Zeitschrift f{\"u}r Physik}\ }\textbf {\bibinfo
  {volume} {9}},\ \bibinfo {pages} {349} (\bibinfo {year} {1922})}\BibitemShut
  {NoStop}%
\bibitem [{\citenamefont {Mott}(1929)}]{mott1929scattering}%
  \BibitemOpen
  \bibfield  {author} {\bibinfo {author} {\bibfnamefont {N.~F.}\ \bibnamefont
  {Mott}},\ }\bibfield  {title} {\bibinfo {title} {The scattering of fast
  electrons by atomic nuclei},\ }\href@noop {} {\bibfield  {journal} {\bibinfo
  {journal} {Proc. R. Soc. London A}\ }\textbf {\bibinfo {volume} {124}},\
  \bibinfo {pages} {425} (\bibinfo {year} {1929})}\BibitemShut {NoStop}%
\bibitem [{\citenamefont {Ando}\ \emph {et~al.}(2011)\citenamefont {Ando},
  \citenamefont {Takahashi}, \citenamefont {Ieda}, \citenamefont {Kajiwara},
  \citenamefont {Nakayama}, \citenamefont {Yoshino}, \citenamefont {Harii},
  \citenamefont {Fujikawa}, \citenamefont {Matsuo}, \citenamefont {Maekawa}
  \emph {et~al.}}]{ando2011inverse}%
  \BibitemOpen
  \bibfield  {author} {\bibinfo {author} {\bibfnamefont {K.}~\bibnamefont
  {Ando}}, \bibinfo {author} {\bibfnamefont {S.}~\bibnamefont {Takahashi}},
  \bibinfo {author} {\bibfnamefont {J.}~\bibnamefont {Ieda}}, \bibinfo {author}
  {\bibfnamefont {Y.}~\bibnamefont {Kajiwara}}, \bibinfo {author}
  {\bibfnamefont {H.}~\bibnamefont {Nakayama}}, \bibinfo {author}
  {\bibfnamefont {T.}~\bibnamefont {Yoshino}}, \bibinfo {author} {\bibfnamefont
  {K.}~\bibnamefont {Harii}}, \bibinfo {author} {\bibfnamefont
  {Y.}~\bibnamefont {Fujikawa}}, \bibinfo {author} {\bibfnamefont
  {M.}~\bibnamefont {Matsuo}}, \bibinfo {author} {\bibfnamefont
  {S.}~\bibnamefont {Maekawa}}, \emph {et~al.},\ }\bibfield  {title} {\bibinfo
  {title} {Inverse spin-{Hall} effect induced by spin pumping in metallic
  system},\ }\href@noop {} {\bibfield  {journal} {\bibinfo  {journal} {J. Appl.
  Phys.}\ }\textbf {\bibinfo {volume} {109}},\ \bibinfo {pages} {103913}
  (\bibinfo {year} {2011})}\BibitemShut {NoStop}%
\bibitem [{\citenamefont {Lindemann}\ \emph {et~al.}(2019)\citenamefont
  {Lindemann}, \citenamefont {Xu}, \citenamefont {Pusch}, \citenamefont
  {Michalzik}, \citenamefont {Hofmann}, \citenamefont {{\v{Z}}uti{\'c}},\ and\
  \citenamefont {Gerhardt}}]{lindemann2019ultrafast}%
  \BibitemOpen
  \bibfield  {author} {\bibinfo {author} {\bibfnamefont {M.}~\bibnamefont
  {Lindemann}}, \bibinfo {author} {\bibfnamefont {G.}~\bibnamefont {Xu}},
  \bibinfo {author} {\bibfnamefont {T.}~\bibnamefont {Pusch}}, \bibinfo
  {author} {\bibfnamefont {R.}~\bibnamefont {Michalzik}}, \bibinfo {author}
  {\bibfnamefont {M.~R.}\ \bibnamefont {Hofmann}}, \bibinfo {author}
  {\bibfnamefont {I.}~\bibnamefont {{\v{Z}}uti{\'c}}},\ and\ \bibinfo {author}
  {\bibfnamefont {N.~C.}\ \bibnamefont {Gerhardt}},\ }\bibfield  {title}
  {\bibinfo {title} {Ultrafast spin-lasers},\ }\href@noop {} {\bibfield
  {journal} {\bibinfo  {journal} {Nature}\ }\textbf {\bibinfo {volume} {568}},\
  \bibinfo {pages} {212} (\bibinfo {year} {2019})}\BibitemShut {NoStop}%
\end{thebibliography}
\end{document}